\documentclass[article, nojss]{jss}\usepackage[]{graphicx}\usepackage{xcolor}
\makeatletter
\def\maxwidth{ %
  \ifdim\Gin@nat@width>\linewidth
    \linewidth
  \else
    \Gin@nat@width
  \fi
}
\makeatother

\usepackage{Sweave}

\usepackage{thumbpdf,lmodern}

\usepackage{float}

\usepackage{subcaption}

\usepackage{booktabs}
\usepackage{tabu}

\author{Nick Barrowman\\CHEO Research Institute\\University of Ottawa
   \And Richard J Webster\\CHEO Research Institute}
\Plainauthor{Nick Barrowman, Richard J Webster}

\title{Exploring data subsets with vtree}
\Plaintitle{Exploring data subsets with vtree}
\Shorttitle{Exploring subsets with vtree}

\Abstract{
\textit{Variable trees} are
a new method for the exploration of discrete multivariate data.
They display nested subsets and
corresponding frequencies and percentages.
Manual calculation of these quantities can be laborious,
especially when there are many multi-level factors and missing data.
Here we introduce variable trees 
and their implementation in the \pkg{vtree} \proglang{R} package,
draw comparisons with existing methods
(contingency tables, mosaic plots, Venn/Euler diagrams, and UpSet),
and illustrate their utility using two case studies.
Variable trees can be used to
(1) reveal patterns in nested subsets,
(2) explore missing data, and
(3) generate study flow diagrams (e.g., CONSORT diagrams) directly from data frames,
to support reproducible research and open science.  
}

\Plainkeywords{Discrete, multivariate, visualization, tree, nested, subsets, R, CONSORT diagram, PRISMA diagram, study flow diagram}

\Address{
  Nick Barrowman\\
  CHEO Research Institute\\
  401 Smyth Road\\
  Ottawa, Ontario, K1H 8L1\\
  \emph{and}\\
  Department of Pediatrics\\
  Faculty of Medicine\\
  University of Ottawa\\
  E-mail: \email{nick.barrowman@gmail.com}\\
  URL: \url{https://nbarrowman.github.io/vtree}\\
\\
  Richard J Webster\\
  CHEO Research Institute\\
  401 Smyth Road\\
  Ottawa, Ontario, K1H 8L1\\
}
\IfFileExists{upquote.sty}{\usepackage{upquote}}{}
\begin{document}



\section[Introduction]{Introduction} \label{sec:intro}

Data exploration is a vital step to gain insights into data sets.
Raw data needs to be cleaned, merged, summarized and assessed.
This process is resource intensive,
accounting for 80\% of time spent on data analysis, by one estimate \citep{Hellerstein2017PrinciplesOD}.
Furthermore, decisions made in this stage can impact scientific rigor and reproducibility.
Recently, an appreciation has emerged
for systematic and transparent protocols about 
data inspection steps to be performed prior to formal data analysis
(e.g., \cite{Huebner2016}).
Such protocols are designed to provide structure at this key stage
while preventing statistical fishing for results.

Tools for data exploration,
like tables and figures,
have been historically important for science.
For instance,
in the late 1800s Florence Nightingale used rose plots to discover
patterns in data that matched her clinical intuition---that most soldiers
in the Crimean War were dying from hygiene-related infections
rather than on the battlefield---and subsequently used this 
to influence the British Parliament \citep{Nelson2012Nightingale}.
This and other methods were a catalyst for the early-1900's
revolution of statistical inference in many scientific fields.

Data exploration tools are more important today than ever.
Data is more ubiquitous with greater volume,
velocity and variety than any time in history \citep{BigData}.
Further, these data are more accessible to analysis due to cheaper
and more powerful computation \citep{waldrop2016chips}.
Consequently,
data literacy and intuitive data exploration tools are required for exploring and communicating findings.  

In this paper we introduce variable trees as a tool for exploring subsets of data, and their implementation in the \pkg{vtree} \proglang{R} package.
The objectives of this paper are
i) to compare variable trees to several established data exploration tools,
ii) to review the functionality of the \pkg{vtree} package, and
iii) to demonstrate the utility of variable trees in two case studies.

\subsection*{Variable trees}

\begin{figure}[!ht]
\centering
\begin{Schunk}

\includegraphics[width=425px]{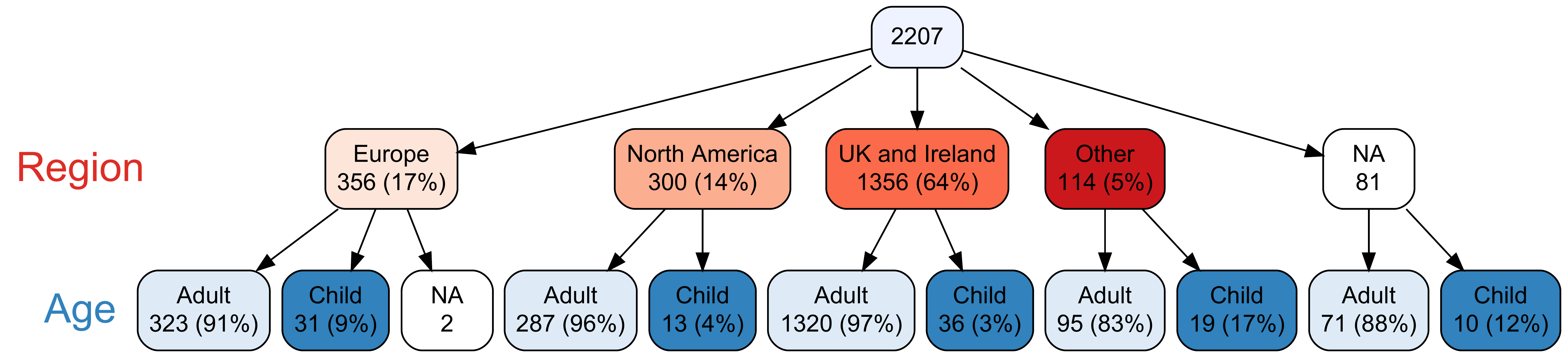} \end{Schunk}
\caption{Variable tree for age nested within region of origin for people onboard the Titanic.}\label{fig:TitanicRegionAge}
\end{figure}

Subsets play an important role in almost any data analysis.
Consider the variables relating to the 2207
passengers and crew members of the Titanic,
represented in the the data set \code{titanic}
from the \pkg{stablelearner} \proglang{R} package \citep{stablelearnerPackage}.
Among other variables,
the data set includes each person's home country,
which we have grouped into regions, and age,
which we have divided into children (under age 13) and adults.
(See the code to perform these transformations at the beginning of
section \ref{sec:PackageFunctionality}.) 
In Figure~\ref{fig:TitanicRegionAge} each person's home region is shown,
and within each region the number and percentage of children and adults are shown.
Missing values are shown as NA.

We call this a \textit{variable tree}.
The \pkg{vtree} package provides a general solution for 
drawing variable trees and describing nested subsets.

Even in simple situations like Figure~\ref{fig:TitanicRegionAge},
it can be a chore to keep track of nested subsets and
calculate the corresponding percentages.
The denominator used to calculate percentages may also depend
on whether the variables have any missing values, as discussed later.
Finally, as the number of variables increases,
the magnitude of the task balloons,
because the number of nested subsets grows exponentially.

\subsection*{The structure of a variable tree}

A variable tree consists of \textit{nodes} connected by arrows.
At the top of Figure~\ref{fig:TitanicRegionAge},
the \textit{root} node of the tree contains
all 2207 people on the Titanic.
The rest of the nodes are arranged in successive layers,
where each layer corresponds to a specific variable.
This highlights one difference between variable trees
and some other kinds of trees:
each layer of a variable tree corresponds to just one variable.
This is distinct from decision trees,
where a layer may include splits based on different variables.
 
The nodes immediately below the root node in Figure~\ref{fig:TitanicRegionAge}
represent values of \code{Region} and
are referred to as the \textit{children}\footnote{Since the Titanic
example involves actual children, we need to be careful here with this technical use of the term ``children''.} of the root node.
Inside each of the nodes, the number of people is displayed
and---except for in a missing value node---the corresponding percentage is also shown.
An example of a missing value node appears in Figure~\ref{fig:TitanicRegionAge},
where \code{Region} was missing (NA)
for 81 people.
Note that, by default, \pkg{vtree} displays ``valid'' percentages,
i.e.,,\ the denominator used to calculate the percentage is the
total number of \textit{non-missing} values,
in this case 2126.
By default,
\pkg{vtree} displays the full missing-value structure of the specified variables.

The final layer of the tree corresponds to values of \code{Age}.
Each of these \textit{leaf} nodes represents children and adults
nested within a subset defined by a value of \code{Region}.
Let's use the notation \code{Region} $\rightarrow$ \code{Age}
to represent \code{Age} nested within \code{Region}.

A missing-value node, like any node, can have children.
For example, of the 81 people for whom \code{Region} is missing,
10 were children and
71 were adults.

\section{Methods of displaying discrete multivariate data} \label{sec:other}

A variety of tools have been developed to display joint distributions of
discrete variables, the most basic being the contingency table,
often enhanced with row, column, or table percentages.
For example,
Table~\ref{tab:contingency1} presents the same information as Figure~\ref{fig:TitanicRegionAge}.
Note that the use of column percentages encourages the reader
to focus on age group nested within region.

\begin{table}[h!]
\centering
\begin{Schunk}

\begin{tabular}{>{}llllll}
\toprule
\multicolumn{1}{l}{\textbf{}} & \multicolumn{1}{l}{\textbf{UK and Ireland}} & \multicolumn{1}{l}{\textbf{Europe}} & \multicolumn{1}{l}{\textbf{North America}} & \multicolumn{1}{l}{\textbf{Other}} & \multicolumn{1}{l}{\textbf{NA}} \\
\cmidrule(l{3pt}r{3pt}){2-2} \cmidrule(l{3pt}r{3pt}){3-3} \cmidrule(l{3pt}r{3pt}){4-4} \cmidrule(l{3pt}r{3pt}){5-5} \cmidrule(l{3pt}r{3pt}){6-6}
 & 1356 (64\%) & 356 (17\%) & 300 (14\%) & 114 (5\%) & 81\\
\midrule
\textbf{Child} & 36 (3\%) & 31 (9\%) & 13 (4\%) & 19 (17\%) & 10 (12\%)\\
\textbf{Adult} & 1320 (97\%) & 323 (91\%) & 287 (96\%) & 95 (83\%) & 71 (88\%)\\
\textbf{NA} & 0 & 2 & 0 & 0 & 0\\
\bottomrule
\end{tabular}

\end{Schunk}
\caption{Contingency table for \code{Region} $\rightarrow$ \code{Age} for people on the Titanic. 
For each combination of region and age,
the frequency and column percentage for age within each region are shown.
Also, directly below the name of each region, the marginal frequency and percentage are shown.
}\label{tab:contingency1}
\end{table}

While the contingency table above is more compact
than the variable tree in Figure~\ref{fig:TitanicRegionAge},
we find the variable tree to be more intuitive.
Furthermore, domain experts often respond well to such
visual representations.

\begin{figure}[H]
\centering
\begin{Schunk}

\includegraphics[width=350px]{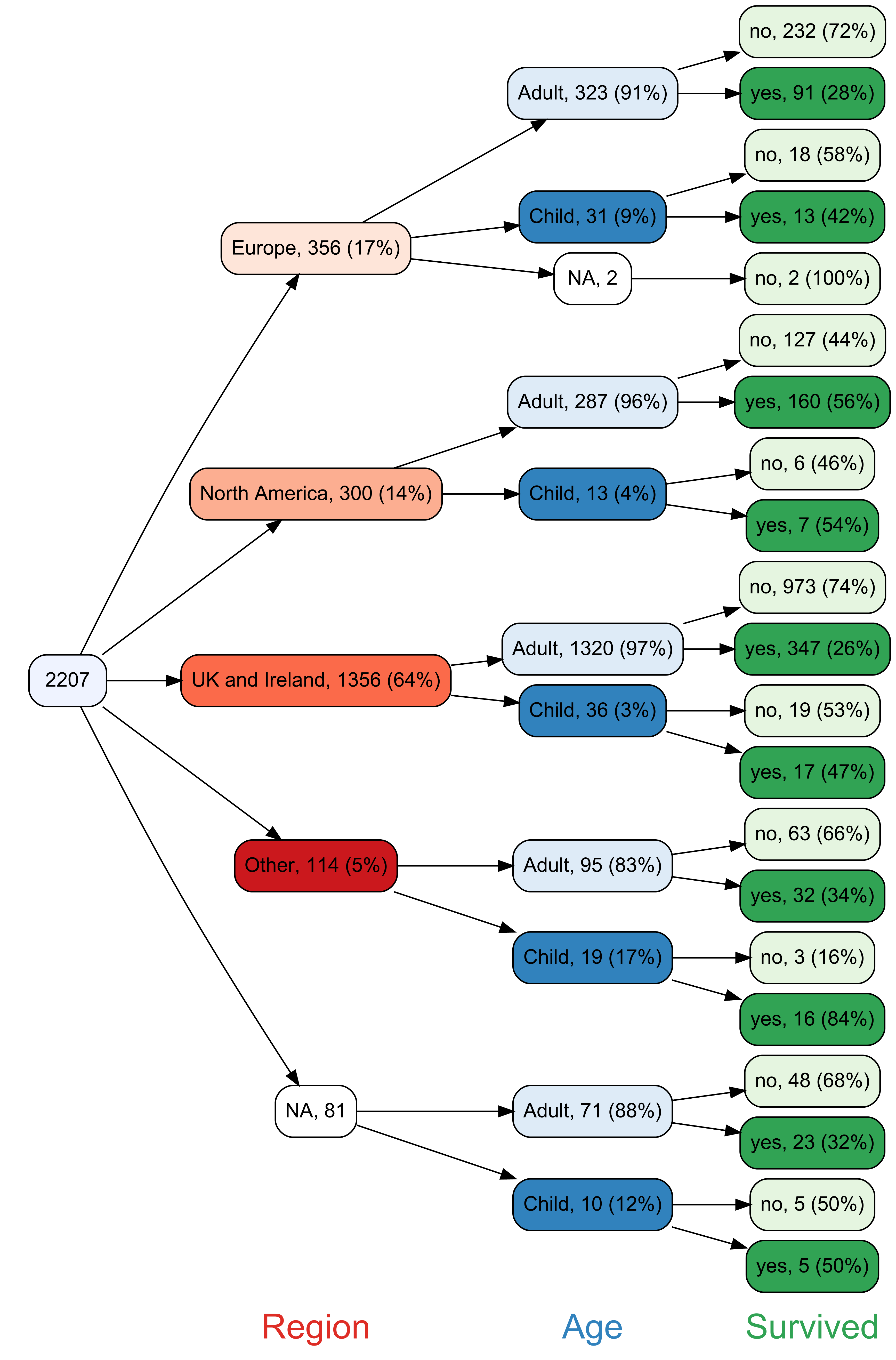} \end{Schunk}
\caption{Variable tree for \code{Region} $\rightarrow$ \code{Age} $\rightarrow$ \code{Survived} with number and percent survival shown in each node. Table~\ref{tab:contingency2} shows the same information.}\label{fig:TitanicRegionAgeSurv}
\end{figure}

Now suppose we'd like to examine
\code{Region} $\rightarrow$ \code{Age} $\rightarrow$ \code{Survived}
(i.e., survival within age within region of origin).
Multi-way cross classifications (three or more variables)
are typically displayed using several two-way tables,
referred to as \textit{layers} or \textit{slices}.
Table~\ref{tab:contingency2} shows two-way tables of 
survival within age group for each of the regions of origin.
This is followed by a variable tree showing the same information
(Figure~\ref{fig:TitanicRegionAgeSurv}).

\begin{table}[h!]
\centering

\vspace{0.2in}
\centerline{\textbf{UK and Ireland} 1356 (64\%)}
\vspace{0.2in}

\begin{Schunk}

\begin{tabular}{>{}l>{\raggedright\arraybackslash}p{0.8in}>{\raggedright\arraybackslash}p{0.8in}>{\raggedright\arraybackslash}p{0.8in}}

\multicolumn{1}{l}{\textbf{}} & \multicolumn{1}{l}{\textbf{Child}} & \multicolumn{1}{l}{\textbf{Adult}} & \multicolumn{1}{l}{\textbf{NA}} \\
 & 36 (3\%) & 1320 (97\%) & 0\\
\midrule
\textbf{Survived} & 17 (47\%) & 347 (26\%) & 0\\
\textbf{Did not survive} & 19 (53\%) & 973 (74\%) & 0\\
\bottomrule
\end{tabular}

\end{Schunk}

\vspace{0.2in}
\centerline{\textbf{North America} 300 (14\%)}
\vspace{0.2in}

\begin{Schunk}

\begin{tabular}{>{}l>{\raggedright\arraybackslash}p{0.8in}>{\raggedright\arraybackslash}p{0.8in}>{\raggedright\arraybackslash}p{0.8in}}

\multicolumn{1}{l}{\textbf{}} & \multicolumn{1}{l}{\textbf{Child}} & \multicolumn{1}{l}{\textbf{Adult}} & \multicolumn{1}{l}{\textbf{NA}} \\
 & 13 (4\%) & 287 (96\%) & 0\\
\midrule
\textbf{Survived} & 7 (54\%) & 160 (56\%) & 0\\
\textbf{Did not survive} & 6 (46\%) & 127 (44\%) & 0\\
\bottomrule
\end{tabular}

\end{Schunk}

\vspace{0.2in}
\centerline{\textbf{Europe} 356 (17\%)}
\vspace{0.2in}

\begin{Schunk}

\begin{tabular}{>{}l>{\raggedright\arraybackslash}p{0.8in}>{\raggedright\arraybackslash}p{0.8in}>{\raggedright\arraybackslash}p{0.8in}}

\multicolumn{1}{l}{\textbf{}} & \multicolumn{1}{l}{\textbf{Child}} & \multicolumn{1}{l}{\textbf{Adult}} & \multicolumn{1}{l}{\textbf{NA}} \\
 & 31 (9\%) & 323 (91\%) & 2\\
\midrule
\textbf{Survived} & 13 (42\%) & 91 (28\%) & 0 (0\%)\\
\textbf{Did not survive} & 18 (58\%) & 232 (72\%) & 2 (100\%)\\
\bottomrule
\end{tabular}

\end{Schunk}

\vspace{0.2in}
\centerline{\textbf{Other} 114 (5\%)}
\vspace{0.2in}

\begin{Schunk}

\begin{tabular}{>{}l>{\raggedright\arraybackslash}p{0.8in}>{\raggedright\arraybackslash}p{0.8in}>{\raggedright\arraybackslash}p{0.8in}}

\multicolumn{1}{l}{\textbf{}} & \multicolumn{1}{l}{\textbf{Child}} & \multicolumn{1}{l}{\textbf{Adult}} & \multicolumn{1}{l}{\textbf{NA}} \\
 & 19 (17\%) & 95 (83\%) & 0\\
\midrule
\textbf{Survived} & 16 (84\%) & 32 (34\%) & 0\\
\textbf{Did not survive} & 3 (16\%) & 63 (66\%) & 0\\
\bottomrule
\end{tabular}

\end{Schunk}

\vspace{0.2in}
\centerline{\textbf{NA} 81}
\vspace{0.2in}

\begin{Schunk}

\begin{tabular}{>{}l>{\raggedright\arraybackslash}p{0.8in}>{\raggedright\arraybackslash}p{0.8in}>{\raggedright\arraybackslash}p{0.8in}}

\multicolumn{1}{l}{\textbf{}} & \multicolumn{1}{l}{\textbf{Child}} & \multicolumn{1}{l}{\textbf{Adult}} & \multicolumn{1}{l}{\textbf{NA}} \\
 & 10 (12\%) & 71 (88\%) & 0\\
\midrule
\textbf{Survived} & 5 (50\%) & 23 (32\%) & 0\\
\textbf{Did not survive} & 5 (50\%) & 48 (68\%) & 0\\
\bottomrule
\end{tabular}

\end{Schunk}

\caption{Contingency table layers for \code{Region} $\rightarrow$ \code{Age} $\rightarrow$ \code{Survived}. The name of each region is shown along with the marginal frequency and percentage, and underneath, the two-way contingency table for \code{Age} $\rightarrow$ \code{Survived} within that region. Along the top row of each table, the marginal frequency and percentage for age within that region is shown. In each table, frequency and column percentage for survival within each age and region are shown.}\label{tab:contingency2}
\end{table}

\clearpage

Note that by default,
\pkg{vtree} shows percentages in each node except for the root.
For example,
of the 2207 people on board the Titanic,
300 (14\%) were from North America,
of whom 287 (96\%) were adults,
of whom 160 (56\%) survived.
In its simplest form, a contingency table only shows crosstabulated frequencies,
corresponding to the frequencies shown in the leaf nodes of a variable tree.
However contingency tables are often shown with
additional marginal and conditional percentages.

As the number of variables increases,
contingency tables for multi-way classifications 
become increasingly difficult to interpret.
In such situations, 
large variable trees can also become unwieldy,
however this can be mitigated by \textit{pruning} branches of lesser interest. 

Contingency tables are not \textit{always} more compact than variable trees.
When most cells of a large contingency table are empty
(in which case the table is said to be \textit{sparse}),
the corresponding variable tree may be
much more compact since empty nodes are not shown.
In the Titanic data set, there are two missing values of Age,
and both are for individuals from Europe. 
This appears as a single node in Figure~\ref{fig:TitanicRegionAgeSurv},
but in Table~\ref{tab:contingency2} in addition to the cell showing 
these 2 missing values, there are 9 cells containing zero.

Like contingency tables,
variable trees show numerical values (frequencies and percentages)
rather than using graphical elements such as area to encode such quantities.
In contrast to contingency tables,
which use a tabular layout to represent subsets,
variable trees use the graphical arrangement of nodes and arrows
to represent the nesting structure.
Put another way, \textit{vtree visualizes relationships, not quantities}.

\subsection*{Visualization of discrete multivariate data}

Several visualization methods have been proposed for discrete multivariate data. Here we focus on displaying raw data and descriptive statistics. Beyond the scope of this paper are various visualizations for fitted models and their residuals \citep{zeileis2007residual}.

Contingency tables can be represented graphically by encoding the quantity in each cell into the size or color of an element.
For example, each cell of a \textit{balloon plot}
contains a circle with area proportional to frequency \citep{moon2017learn}.
Balloon plots can be produced in R using the function \code{ggballoonplot} in the \pkg{ggpubr} package \citep{ggpubrPackage}.
Concordance of ordinal categorical data may be visualized using \textit{agreement charts} \citep{bangdiwala2013agreement};
these can be produced in R using the \code{agreementplot} function from the \pkg{vcd} package \citep{hornik2006strucplot}. 

Barplots are another common discrete multivariate data visualization tool.
They can also be produced for subsets, deﬁned by values of another variable.
A more compact representation is the stacked barplot and grouped stacked barplot.
A limitation with stacked barplots is the difficulty of between-category comparisons,
since there is no common baseline, except for the bottom category in a stack. 

\begin{figure}[H]
\centering
\begin{Schunk}

\includegraphics[width=\maxwidth]{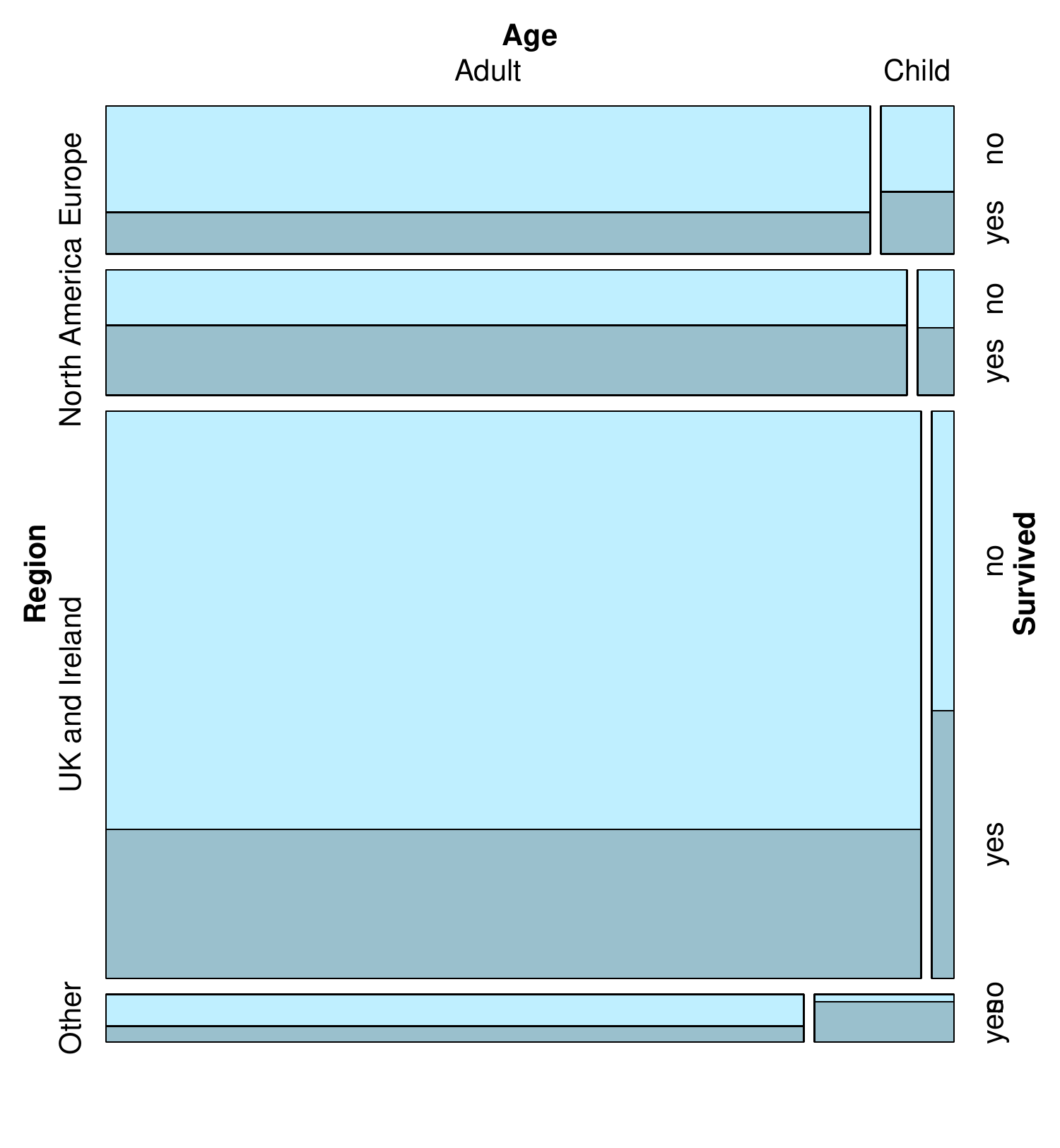} \end{Schunk}
\caption{Mosaic plot for \code{Region} $\rightarrow$ \code{Age} $\rightarrow$ \code{Survived} for people on the Titanic. Each rectangle corresponds to a subset of the data and the area of the rectangle represents the relative frequency. Table~\ref{tab:contingency2} and Figure~\ref{fig:TitanicRegionAgeSurv} show the same information.}\label{fig:MosaicFig}
\end{figure}

An elegant extension of the stacked barplot is the \textit{mosaic plot} \citep{HartiganKleiner:1981}.
In a mosaic plot, the area of each rectangle represents the number of observations in the corresponding subset of the data.
Mosaic plots are available in R through several packages
(e.g., \pkg{vcd}, \pkg{ggmosaic} \citep{ggmosaicPackage}, \pkg{iplots} \citep{iplotsPackage}, as well as in the base R function \code{mosaicplot}. The Strucplot framework implemented in the \pkg{vcd} package provides flexibility for hierarchical conditional plots (e.g., conditional mosaic plots, association plots, double decker plots, sieve plots and more).
Mosaic plots can provide an intuitive visual representation of the number of observations in subsets of the data, however they tend to become visually complicated when there are more than three variables.
Figure~\ref{fig:MosaicFig} is a mosaic plot for 
\code{Region} $\rightarrow$ \code{Age} $\rightarrow$ \code{Survived}
for the people onboard the Titanic,
as in Table~\ref{tab:contingency2} and Figure~\ref{fig:TitanicRegionAgeSurv}.

\clearpage

Visualizations like Figure~\ref{fig:MosaicFig} have advantages and disadvantages compared to text and tabular summaries.
On the one hand, they represent quantitative and qualitative information
in a way that is quickly decoded by our visual perceptual systems.
On the other, visualizations can be unfamiliar and even perplexing compared
to numerical and tabular representations.
A practical advantage of tabular summaries is that they are more easily formatted and manipulated using current software.
Variable trees share characteristics
with both tabular representations and visualizations.
To put it succinctly,
variable trees visualize \textit{relationships} not \textit{quantities}.

\subsection*{Data representing set membership}

\begin{figure}[H]
\centering
\begin{Schunk}

\includegraphics[width=\maxwidth]{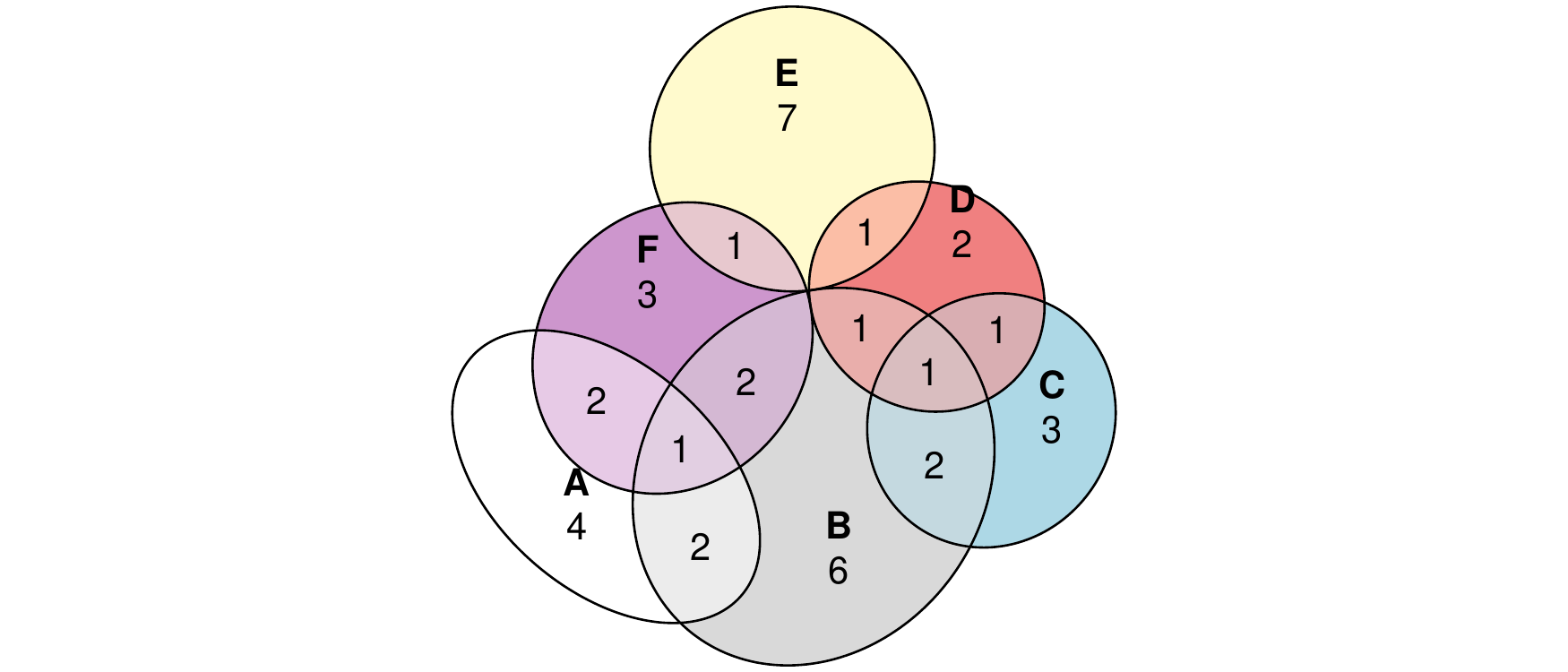} \end{Schunk}
\caption{Euler plot for the dataset of \cite{Wilkinson:2012}.}\label{fig:Euler}
\end{figure}

A special type of discrete multivariate data is when all of the variables are binary,
in which case they can be interpreted as representing set membership.
Venn and Euler diagrams have long been used to represent the intersection of sets.
Venn diagrams use overlapping closed curves such that all intersections between sets
are represented by overlapping areas.
Euler diagrams are like Venn diagrams but empty intersections need not be shown.
For data sets, software is available to calculate
the number of observations in each of the intersections,
for example in R, the \pkg{VennDiagram} \citep{VennDiagramPackage} and
\pkg{venneuler} \citep{venneulerPackage} packages.
A further elaboration of these diagrams is to make the areas of the sets and 
their intersections approximately proportional to the number of observations in each subset.
The package \pkg{eulerr} \citep{eurlerrPackage} provides this functionality.
For example, in Figure~\ref{fig:Euler} a dataset of \cite{Wilkinson:2012} is
represented using an approximately area-proportional Euler diagram.
As the number of sets grows,
Venn and Euler diagrams can become unwieldy.

\begin{figure}[H]
\centering
\begin{Schunk}

\includegraphics[width=\maxwidth]{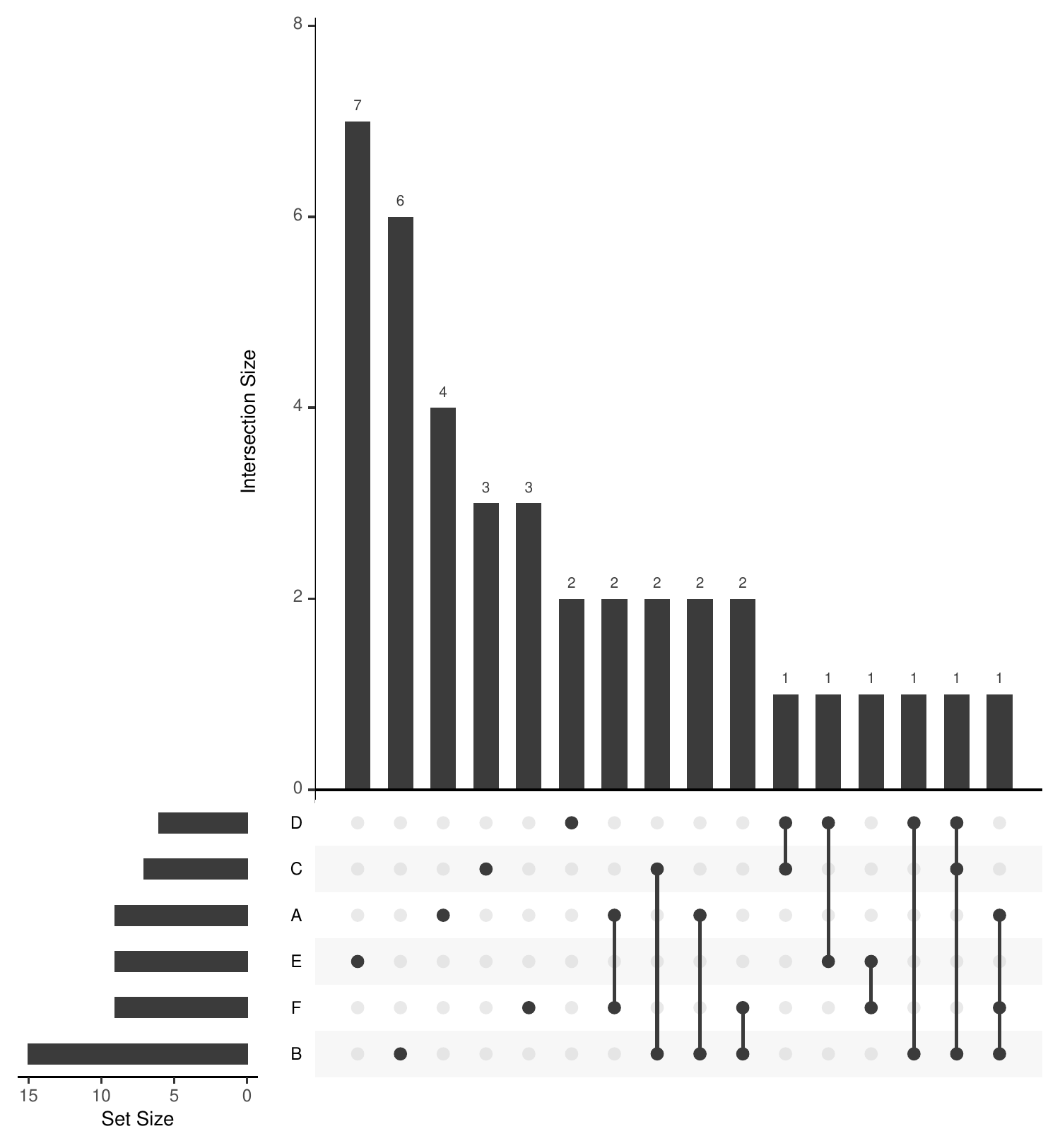} \end{Schunk}
\caption{An UpSet plot}\label{fig:UpSet}
\end{figure}

An innovative way to represent the intersections of a large number of sets is the \textit{UpSet} plot
\citep{Lex:2014}.
The R package \pkg{UpSetR} \citep{Conway:2017} was used to produce Figure~\ref{fig:UpSet}
for the dataset of \cite{Wilkinson:2012}.
UpSet uses a grid layout to represent the intersections
(see the dots at the bottom of Figure~\ref{fig:UpSet}),
together with bar graphs on each side to represent the size of sets and intersections.

\begin{figure}[H]
\centering
\begin{Schunk}

\includegraphics[width=350px]{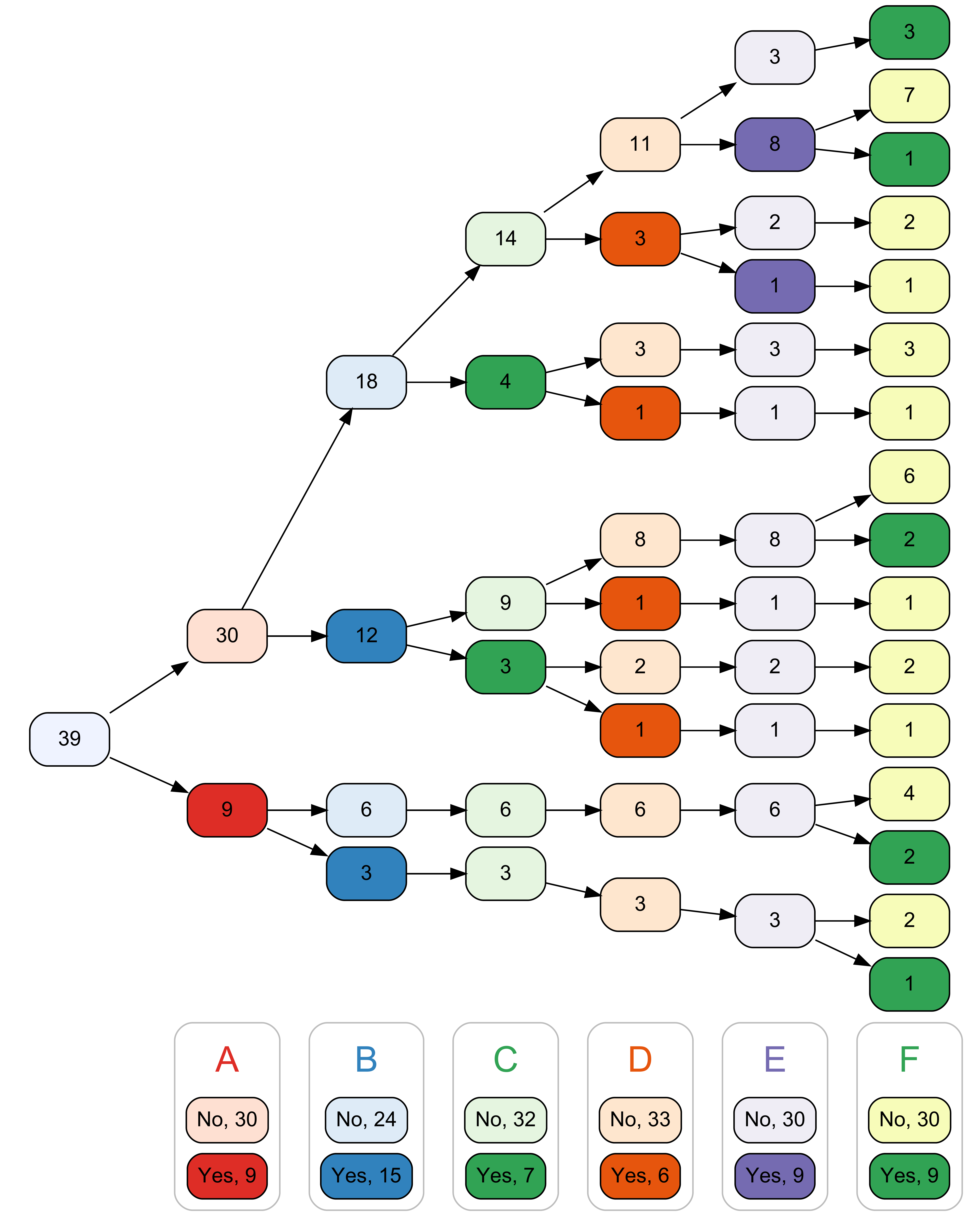} \end{Schunk}
\caption{A variable tree for the dataset of \cite{Wilkinson:2012}. Note that \textit{legend nodes} below the tree show the marginal frequencies for each variable.}\label{fig:VennTree}
\end{figure}

Variable trees can also represent the intersection of sets,
however unlike UpSet and area-proportional Euler diagrams,
they do not use graphical elements to encode quantity.
Like non-proportional Venn Diagrams,
variable trees graphically depict the relationships between subsets of the data,
but represent quantities numerically (Figure~\ref{fig:VennTree}).
Unlike Venn, Euler, and UpSet diagrams,
variable trees require a prespecified ordering.
For example,
Figure~\ref{fig:VennTree} uses the ordering A $\rightarrow$ B $\rightarrow$ C $\rightarrow$ D $\rightarrow$ E $\rightarrow$ F.

\begin{figure}[H]
\centering
\begin{Schunk}

\includegraphics[width=400px]{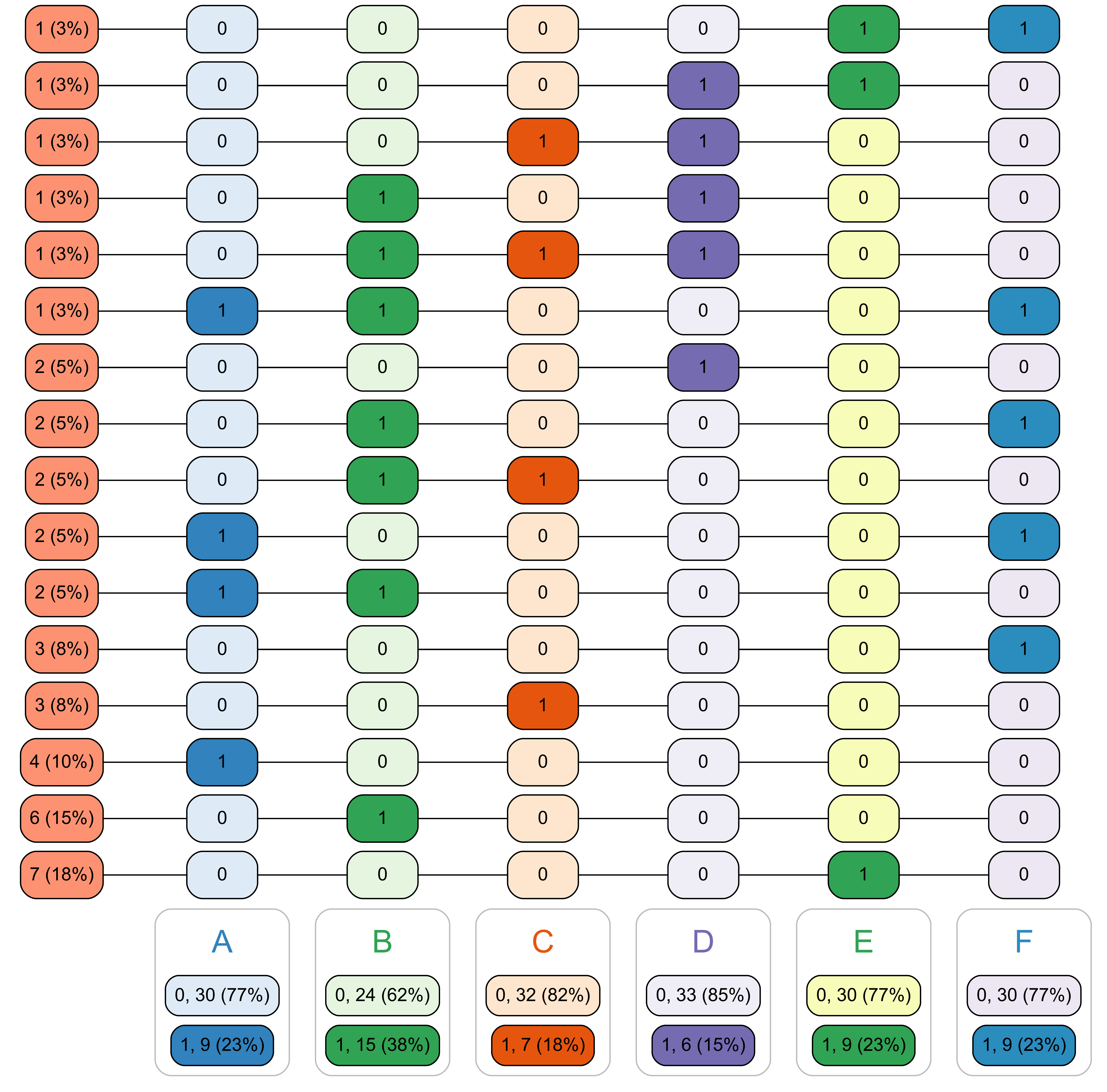} \end{Schunk}
\caption{A pattern tree for the dataset of \cite{Wilkinson:2012}. Each row represents a "pattern"
(here a particular intersection), corresponding to a terminal node in Figure~\ref{fig:VennTree}.}\label{fig:VennPattern}
\end{figure}

An alternative diagram that \pkg{vtree} can produce is the \textit{pattern tree},
which depicts every intersection.
Each row in Figure~\ref{fig:VennPattern} corresponds to the combination of
values represented by a terminal node in Figure~\ref{fig:VennTree}.
Since the intermediate nodes in Figure~\ref{fig:VennTree} are not represented,
information is lost.
The pattern tree is much easier to read, however.
Pattern trees have some of the same structure as an UpSet plot,
except that sizes of subsets are not represented graphically as in 
the bar graphs on the sides of an UpSet plot.

\section{Package functionality} \label{sec:PackageFunctionality}

This section provides an overview of the features of the \pkg{vtree} package.
Additional resources are available in
the package 
\href{https://cran.r-project.org/web/packages/vtree/vignettes/vtree.html}{vignette}, 
a \href{https://nbarrowman.github.io/cheatsheets/vtree_cheatsheet_5.0.0.pdf}{cheatsheet},
and \href{https://www.youtube.com/playlist?list=PLB_c7T0wmWrotNHViouASt1VO_DOH9J5J}{video tutorials} on YouTube.

To illustrate the functionality of the \pkg{vtree} package,
we use the \code{titanic} data set from the \pkg{stablelearner} package.

\begin{Schunk}
\begin{Sinput}
R> library(vtree)
R> library(stablelearner)
R> library(dplyr)
R> library(forcats)
R> data("titanic")
R> td <- titanic %>%
+   rename(Survived = survived) %>%
+   mutate(
+     Age = ifelse(age<13, "Child", "Adult"),
+     Gender = gender,
+     Class = fct_collapse(class,
+       Crew = c("deck crew", "engineering crew", "restaurant staff",
+         "victualling crew")),
+     Region = fct_collapse(country,
+       "UK and Ireland" = c("England", "Scotland", "Wales", "Ireland",
+         "Northern Ireland", "Channel Islands"),
+       "Europe" = c("Norway", "France", "Finland", "Sweden", "Latvia",
+         "Denmark", "Bulgaria", "Greece", "Hungary", "France", "Spain", 
+         "Italy", "Belgium", "Germany", "Austria", "Poland", 
+         "Switzerland", "Bosnia", "Croatia", "Croatia (Modern)", 
+         "Yugoslavia", "Slovakia (Modern day)", "Slovenia", "Netherlands",
+         "Russia"),
+       "North America" = c("United States", "Canada", "Mexico", "Cuba"),
+       other_level = "Other"))
\end{Sinput}
\end{Schunk}

\subsection*{Calling vtree}

Suppose the Titanic data are in a data frame called \code{td}.
To display a variable tree for a single variable, say \code{Class}, 
use the following command:

\begin{figure}[H]
\centering
\begin{Schunk}
\begin{Sinput}
R> vtree(td, "Class")
\end{Sinput}

\includegraphics[width=120px]{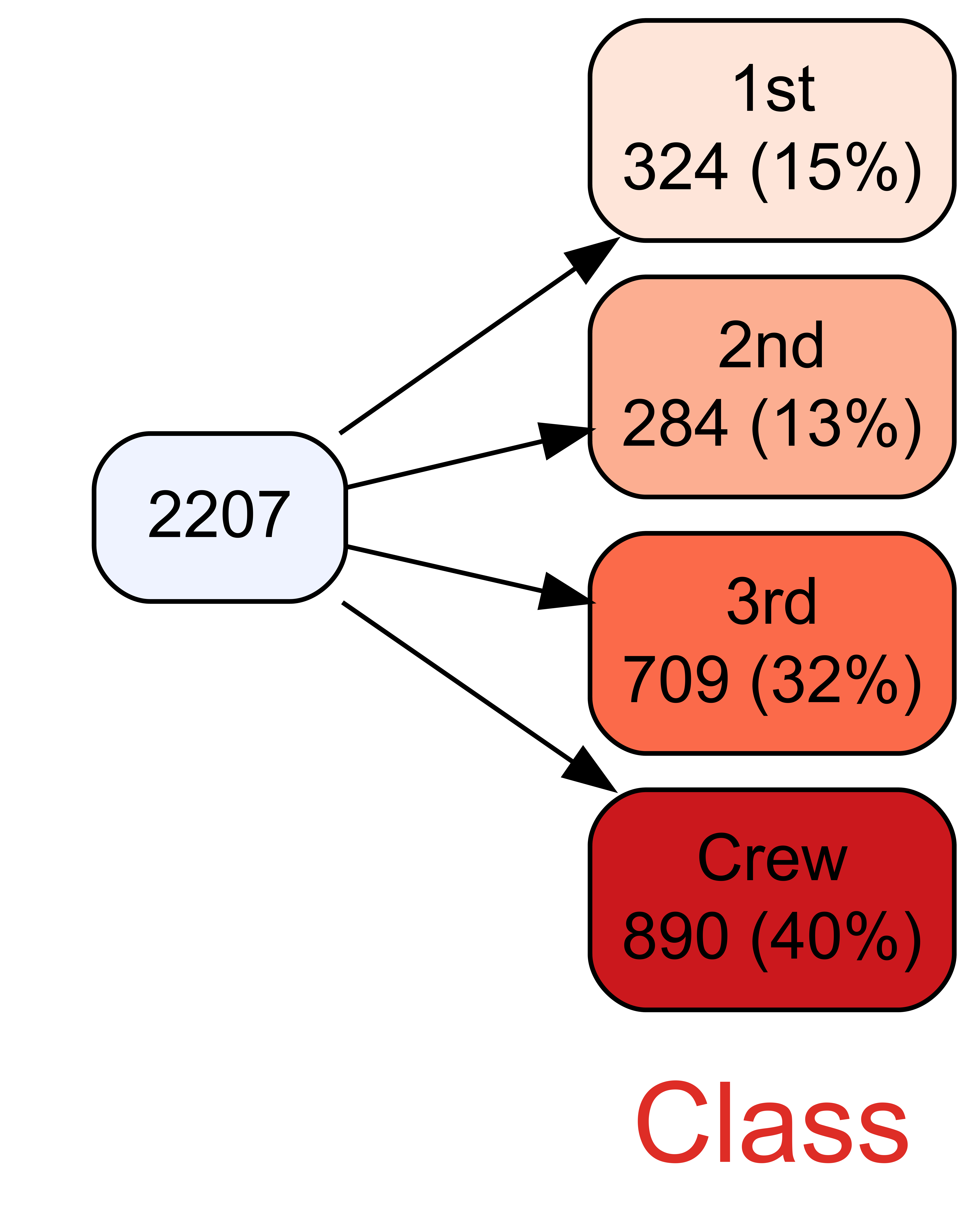} \end{Schunk}
\caption{A simple variable tree.}\label{fig:simplevtree}
\end{figure}

To produce a variable tree for \code{Class} $\rightarrow$ \code{Age},
specify \code{"Class Age"}:

\begin{figure}[H]
\centering
\begin{Schunk}
\begin{Sinput}
R> vtree(td, "Class Age", horiz = FALSE)
\end{Sinput}

\includegraphics[width=400px]{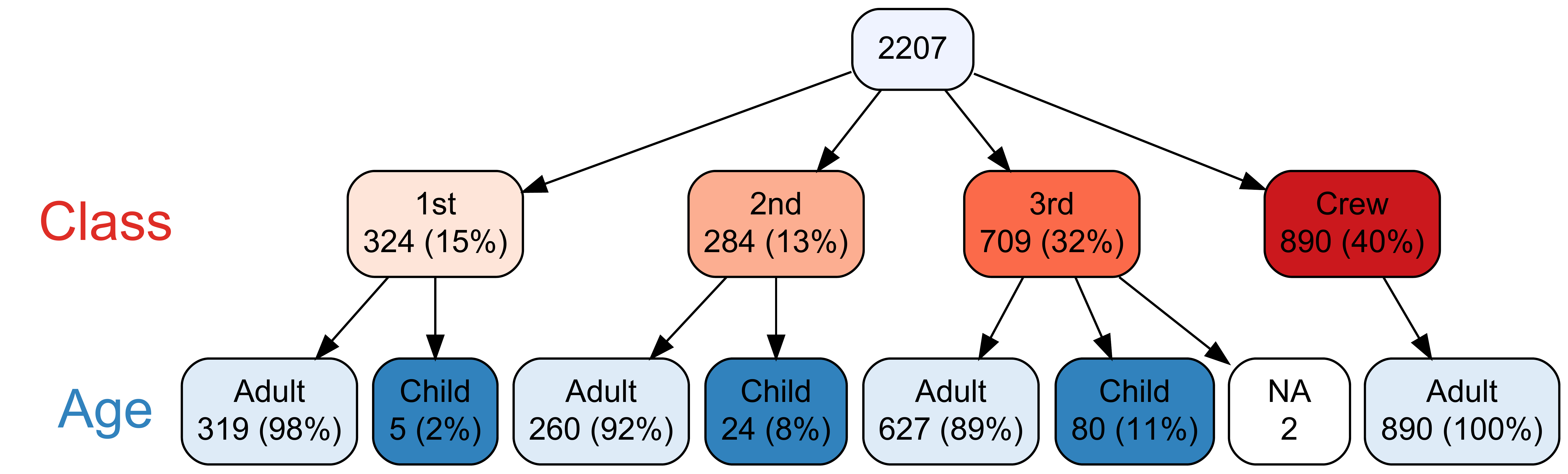} \end{Schunk}
\caption{A two-layer vertical variable tree.}\label{fig:simplevtree2}
\end{figure}

By default, vtree produces horizontal trees.
The tree in Figure~\ref{fig:simplevtree2} is vertical because of
the specification \code{horiz = FALSE}.

Note that \code{vtree} can also be called at the end of a \pkg{dplyr} pipeline.
This allows variables and data to be modified for use in a variable tree.
For example, the following commands produce
a tree like the one in Figure~\ref{fig:simplevtree2},
but \textit{only} including female passengers and crew:

\begin{figure}[H]
\centering
\begin{Schunk}
\begin{Sinput}
R> library(dplyr)
R> td %>% filter(Gender == "female") %>% vtree(~ Class + Age, horiz = FALSE)
\end{Sinput}

\includegraphics[width=400px]{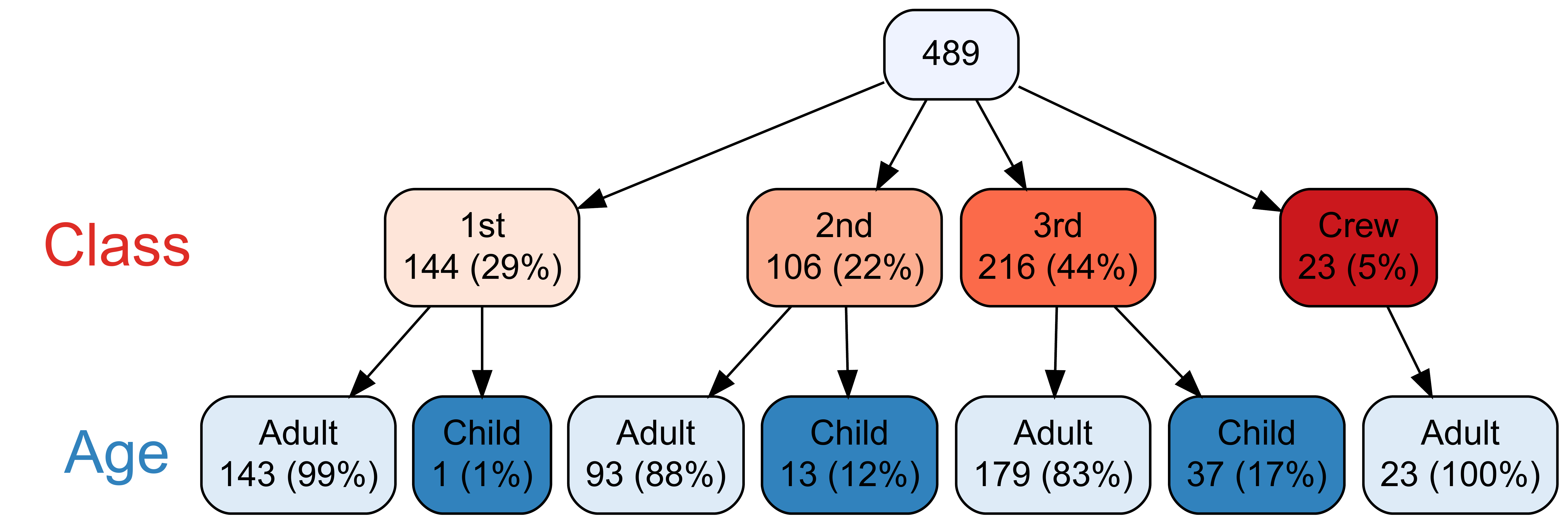} \end{Schunk}
\caption{Class of female passengers and crew onboard the Titanic.}\label{fig:simplevtree2_female}
\end{figure}
 
The example above includes the call \code{vtree(~ Class + Age, horiz = FALSE)},
which uses a formula to specify variables rather than a character string.
(This notation is convenient, 
however it precludes the use of \pkg{vtree}'s
built-in \hyperref[sec:VarSpec]{variable specifications}.)

\subsection*{Pruning}

When a variable tree gets too big,
or you are only interested in certain parts of the tree,
it may be useful to remove some nodes along with their descendants.
This is known as \textit{pruning}.
For convenience, there are several different ways to prune a tree,
described below.

Suppose you don't wish to show the ``Europe'' node or the ``Other'' node 
(which represents people from other parts of the world such
as India, the Middle East, etc.).
Specifying \code{prune=list(Region = c("Europe", "Other"))}
removes those nodes, and all of their descendants:

\begin{figure}[H]
\centering
\begin{Schunk}
\begin{Sinput}
R> vtree(td, "Region Age", prune = list(Region = c("Europe", "Other")), 
+   horiz = FALSE)
\end{Sinput}

\includegraphics[width=325px]{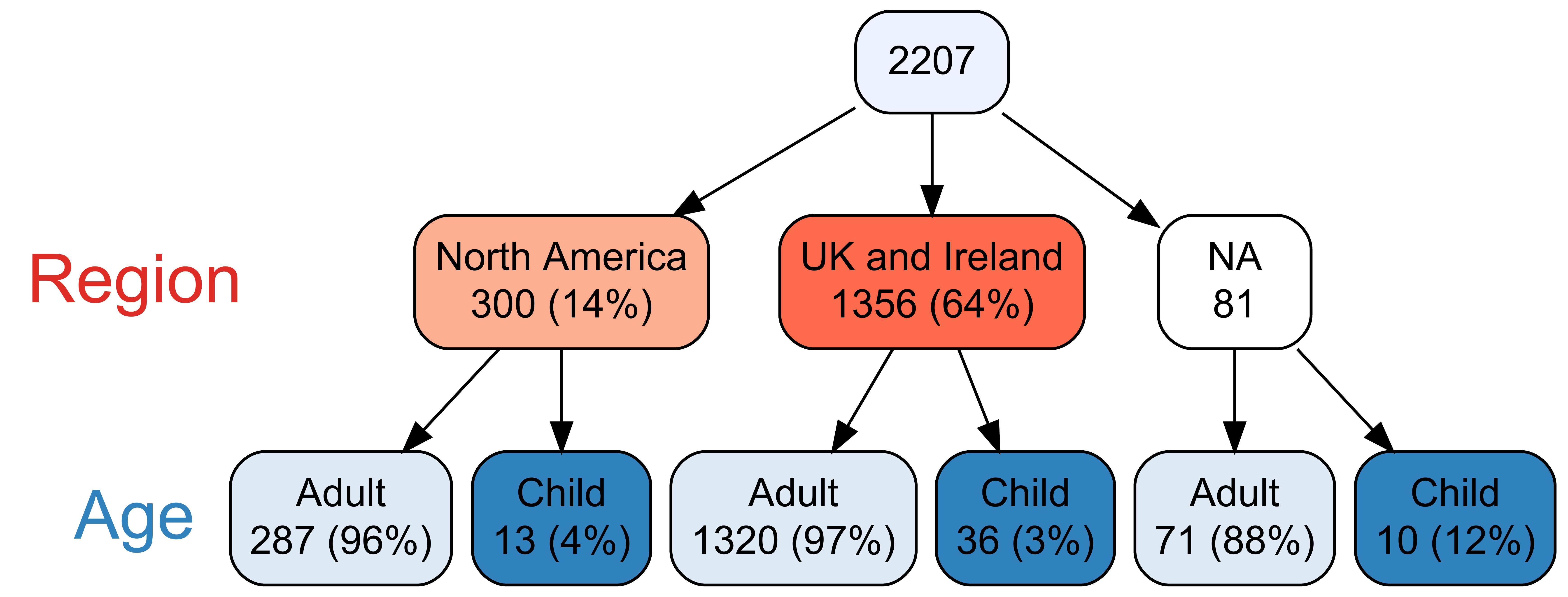} \end{Schunk}
\caption{\label{fig:prune1}Using the prune parameter to remove ``Europe'' and ``Other'' regions.}
\end{figure}

In general,
the argument of the \code{prune} parameter is a \textit{list}
with an element named for each variable you wish to prune.
In the example above, the list has a single element, named \code{Region}.
In turn, that element is a vector \code{c("Europe", "Other")}
indicating the values of \code{Region} to prune.

Note that once a variable tree has been pruned, it is no longer complete.
This can sometimes be confusing since not all observations
are represented at certain layers of the tree.
For example in the tree above, only 
1737 observations
are shown in the \code{Region} nodes and their children.

Sometimes it is more convenient to specify which nodes should be \textit{retained}
rather than which ones should be discarded.
The \code{keep} parameter is used for this purpose,
and can thus be considered the complement of the \code{prune} parameter.
Suppose we wish to create a variable tree for
\code{Region} $\rightarrow$  \code{Class} $\rightarrow$ \code{Gender} $\rightarrow$ \code{Age}.
This tree has four layers, and without any pruning it would be quite large.
But suppose we are only interested in certain branches of the tree,
say 
the ``Europe'' node of \code{Region},
the ``3rd'' node of \code{Class},
and
the ``male'' node of \code{Gender}.
Using the \code{keep} parameter, a compact tree can be produced:

\begin{figure}[H]
\centering
\begin{Schunk}
\begin{Sinput}
R> vtree(td, "Region Class Gender Age", 
+   keep = list(Region = "Europe", Class = "3rd", Gender = "male"))
\end{Sinput}

\includegraphics[width=300px]{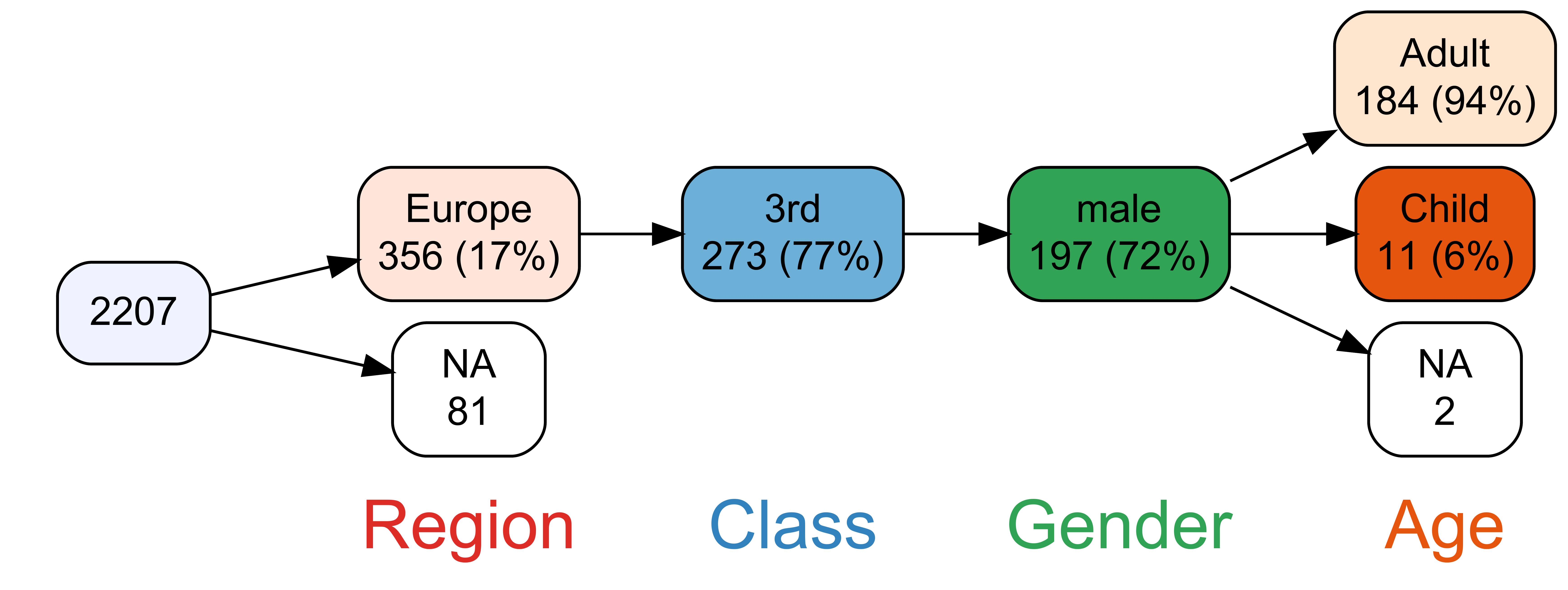} \end{Schunk}
\caption{Using the keep parameter.}\label{fig:keepEurope}
\end{figure}

In the \code{Region} layer of Figure~\ref{fig:keepEurope},
note that along with the ``Europe'' node,
the missing value node has also been retained.
In general, whenever valid percentages are used (which is the default),
missing value nodes are retained when \code{keep} is used.
This is because valid percentages are difficult to interpret without
knowing the denominator, which requires knowing the number of missing values.
On the other hand, here's what happens when \code{vp = FALSE}:

\begin{figure}[H]
\centering
\begin{Schunk}
\begin{Sinput}
R> vtree(td, "Region Class Gender Age",
+   keep = list(Region = "Europe", Class = "3rd", Gender = "male"),
+   vp = FALSE)
\end{Sinput}

\includegraphics[width=300px]{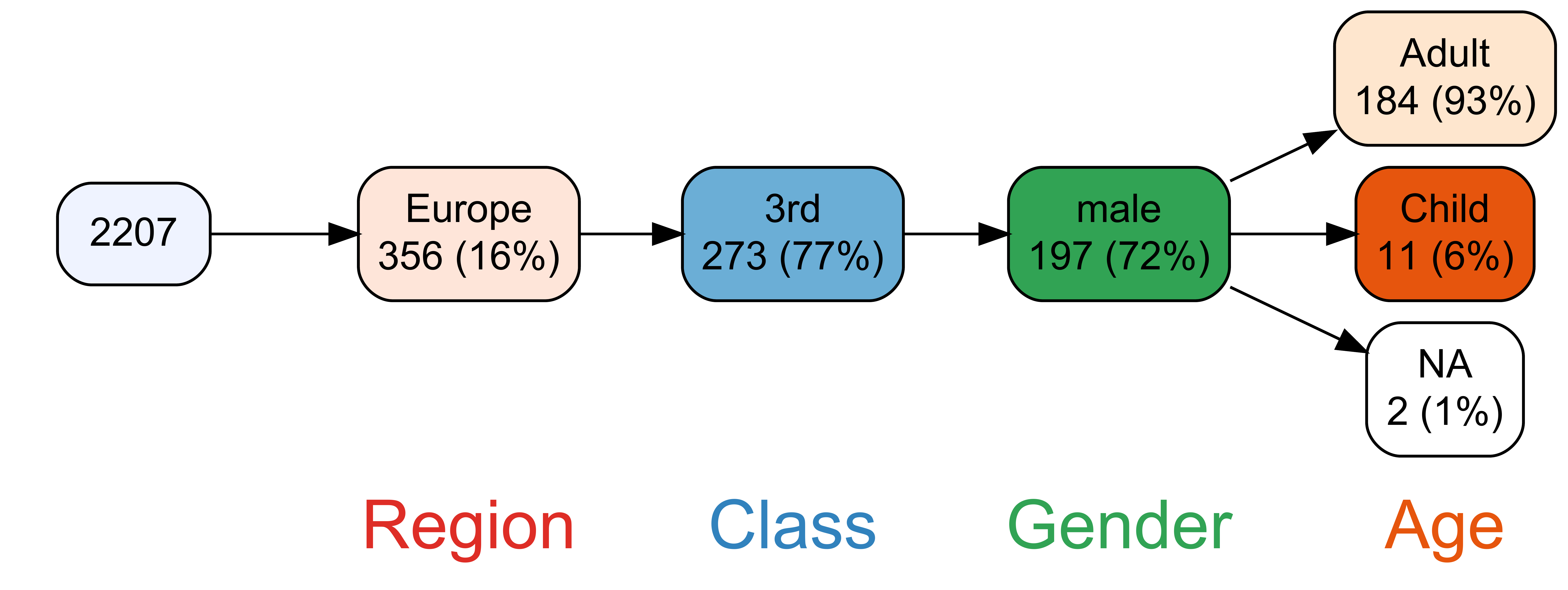} \end{Schunk}
\caption{Same as Figure~\ref{fig:keepEurope} but not using ``valid'' percentages.}\label{fig:fourlevels}
\end{figure}

Note that the missing value node for \code{Region} is no longer present,
since the percentage for the ``Europe'' node can be interpreted without
knowing the number of missing values.
Also, note that the missing value node for \code{Age} includes a percentage,
and the percentages for the other nodes of \code{Age} are slightly different.
(With only two missing values, the difference is slight,
but as the proportion of missing data increases,
the percentages become substantially different.)

An alternative is to prune \textit{below} the specified nodes
(i.e., to prune their descendants), so that the counts always add up.
In the present example, this means that the other nodes will be shown,
but not their descendants.
The \code{prunebelow} parameter is used to do this:

\begin{figure}[H]
\centering
\begin{Schunk}
\begin{Sinput}
R> vtree(td, "Region Age",
+   prunebelow = list(Region =c("UK and Ireland", "North America", "Other")))
\end{Sinput}

\includegraphics[width=175px]{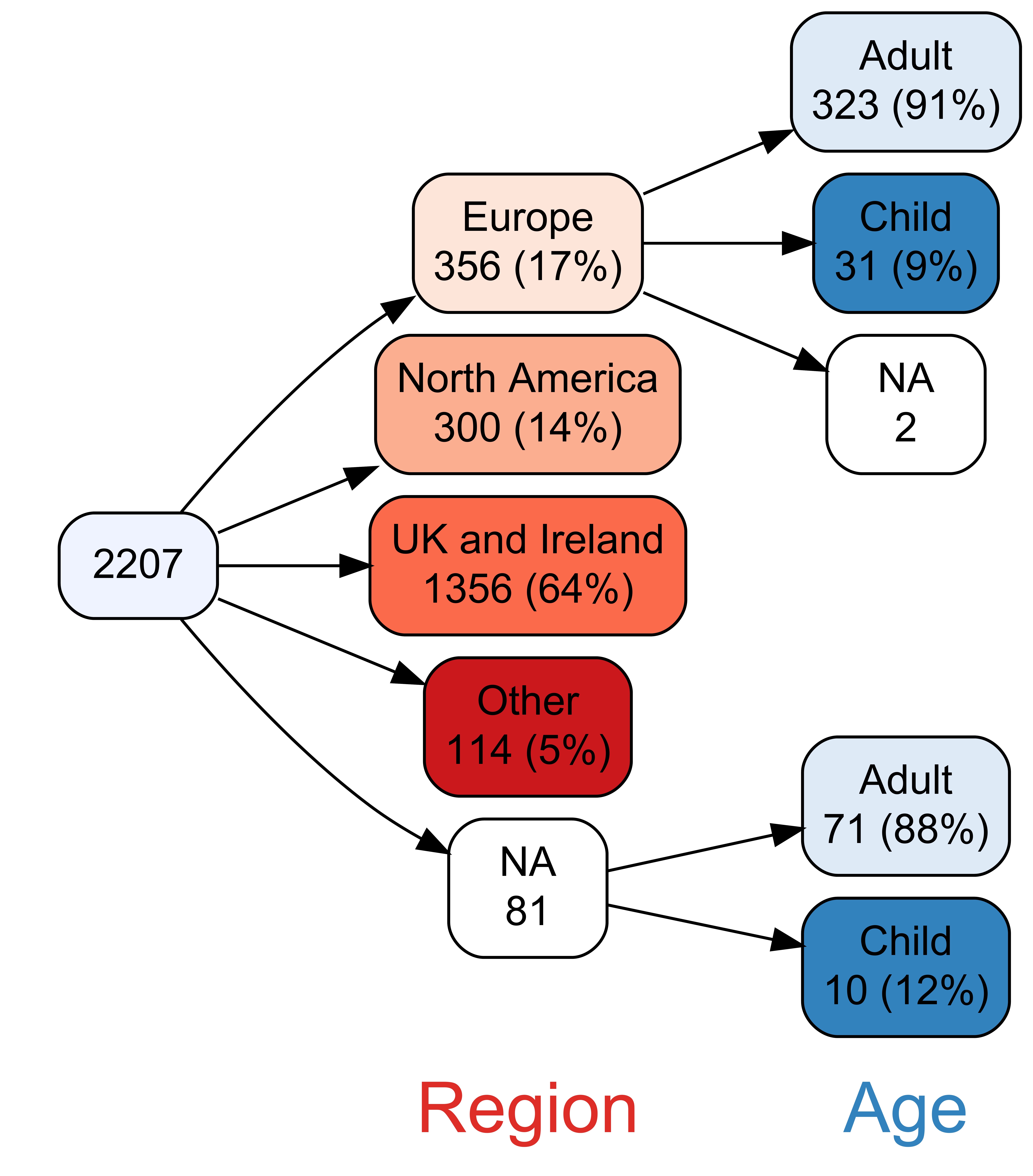} \end{Schunk}
\caption{Using the prunebelow parameter.}\label{fig:prunebelow}
\end{figure}

The complement of the \code{prunebelow} parameter is the \code{follow} parameter.
Instead of specifying which nodes should be pruned below,
this allows you to specify which nodes should be \textit{followed}
(that is, \textit{not} pruned below).

As a variable trees grow,
it can become difficult to see the forest for the tree, as it were.
For example, the following variable tree is very wide,
which makes it difficult to read.
(One small modification to make it a little narrower
is the use of the \code{splitwidth} parameter,
which specifies the number of characters of text
before it gets split onto another line.)

\begin{figure}[H]
\centering
\begin{Schunk}
\begin{Sinput}
R> vtree(td, "Class Region", horiz = FALSE, splitwidth = 5)
\end{Sinput}

\includegraphics[width=435px]{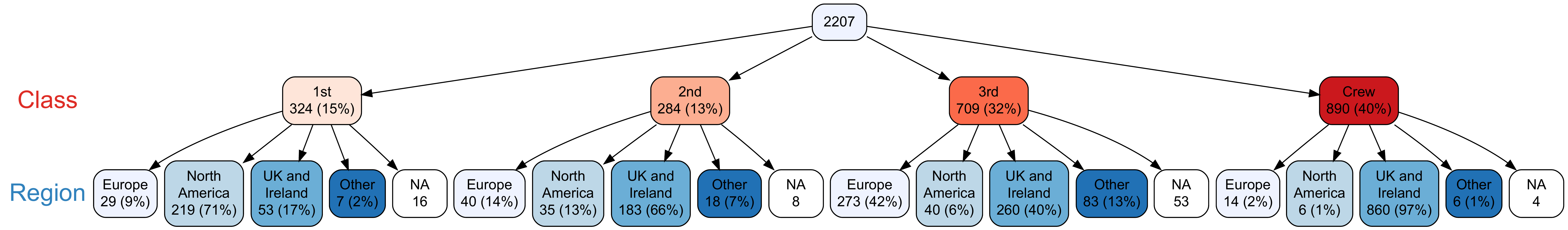} \end{Schunk}
\caption{A variable tree that is too wide to read easily.}\label{fig:hardtoread}
\end{figure}

One solution is to prune nodes that contain small numbers of observations.
For example if you want to only see nodes with at least 50 observations,
you can specify \code{prunesmaller = 50}, as in this example:

\begin{figure}[H]
\centering
\begin{Schunk}
\begin{Sinput}
R> vtree(td, "Class Region", horiz = FALSE, prunesmaller = 50,
+   splitwidth = 5)
\end{Sinput}

\includegraphics[width=435px]{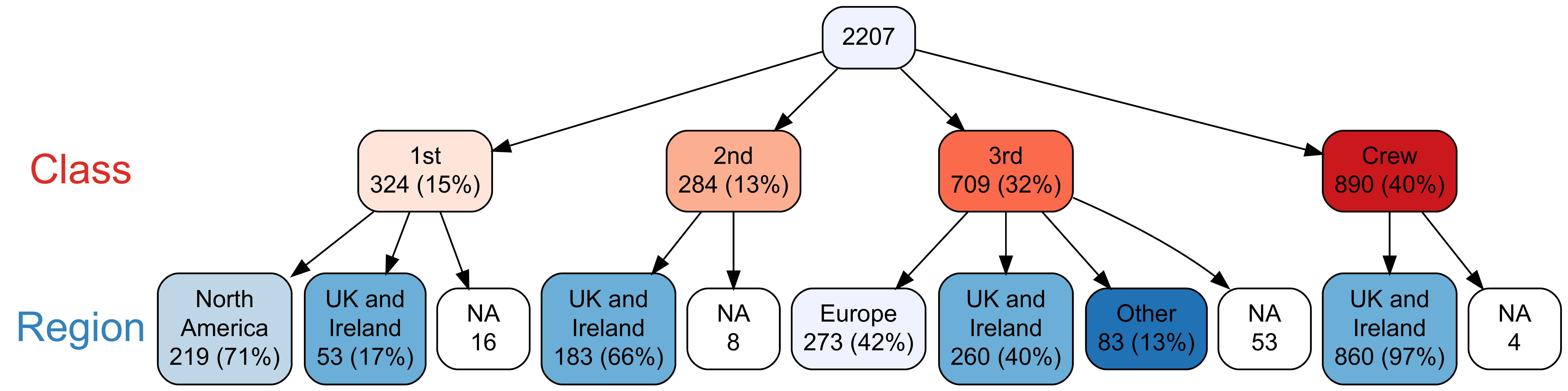} \end{Schunk}
\caption{Using the prunesmaller parameter.}\label{fig:prunesmaller}
\end{figure}

Similar to the \code{keep} parameter,
when valid percentages are used (\code{vp = TRUE}, which is the default), 
nodes represent missing values will not be pruned.
As noted previously,
this is because percentages are confusing when missing values are not shown.
On the other hand,
when \code{vp = FALSE}, missing nodes can be pruned.

\subsection*{Labels for variables and nodes}

Readability of a variable tree can be improved 
by customizing the variable and node names
using the \code{labelvar} and \code{labelnode} parameters.
By default, \pkg{vtree} labels variables and nodes
exactly as they appear in the data frame.

For example, the \code{embarked} variable indicates the port where
a passenger or crew member boarded the Titanic.
Suppose we wish this variable to appear as \textit{Port} in the 
variable tree.
The \code{labelvar} parameter specifies this.

\newpage

\begin{figure}[H]
\centering
\begin{Schunk}
\begin{Sinput}
R> vtree(td, "Class embarked", labelvar = c("embarked" = "Port"),
+   sameline = TRUE)
\end{Sinput}

\includegraphics[width=250px]{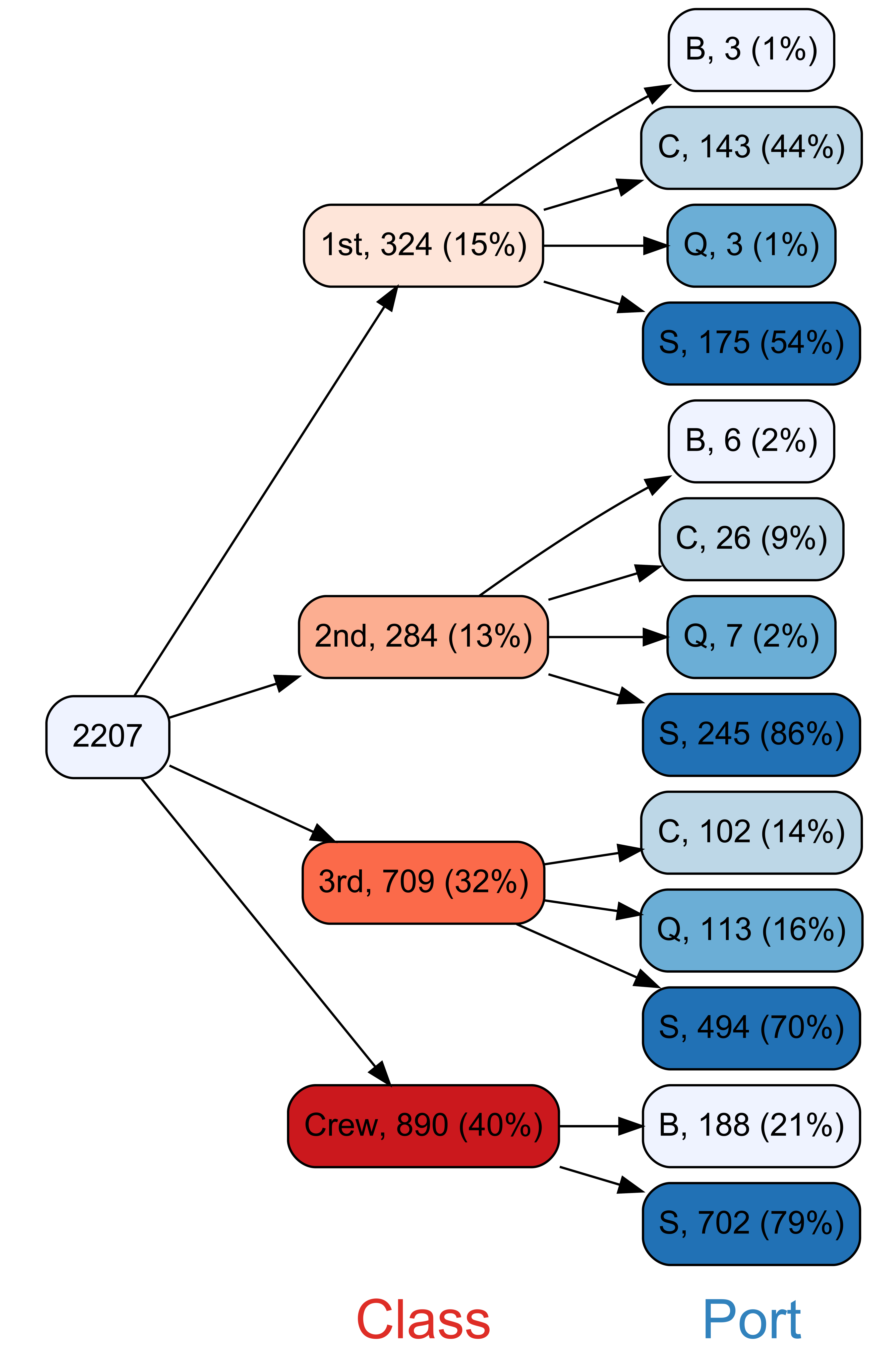} \end{Schunk}
\caption{Using the labelvar parameter.}\label{fig:labelvar}
\end{figure}

By default, \code{vtree} labels nodes (except for the root node)
using the values of the variable in question.
(If the variable is a factor, the levels of the factor are used).
Sometimes it is convenient to instead specify custom labels for nodes.
The \code{labelnode} argument can be used to relabel the values.
For example, to relabel the classes as
``First Class'', ``Second Class'', and ``Third Class'':

\begin{figure}[H]
\centering
\begin{Schunk}
\begin{Sinput}
R> vtree(td, "Class", horiz = FALSE, labelnode = list(Class = c(
+   "First Class" = "1st", "Second Class" = "2nd", "Third Class" = "3rd")))
\end{Sinput}

\includegraphics[width=250px]{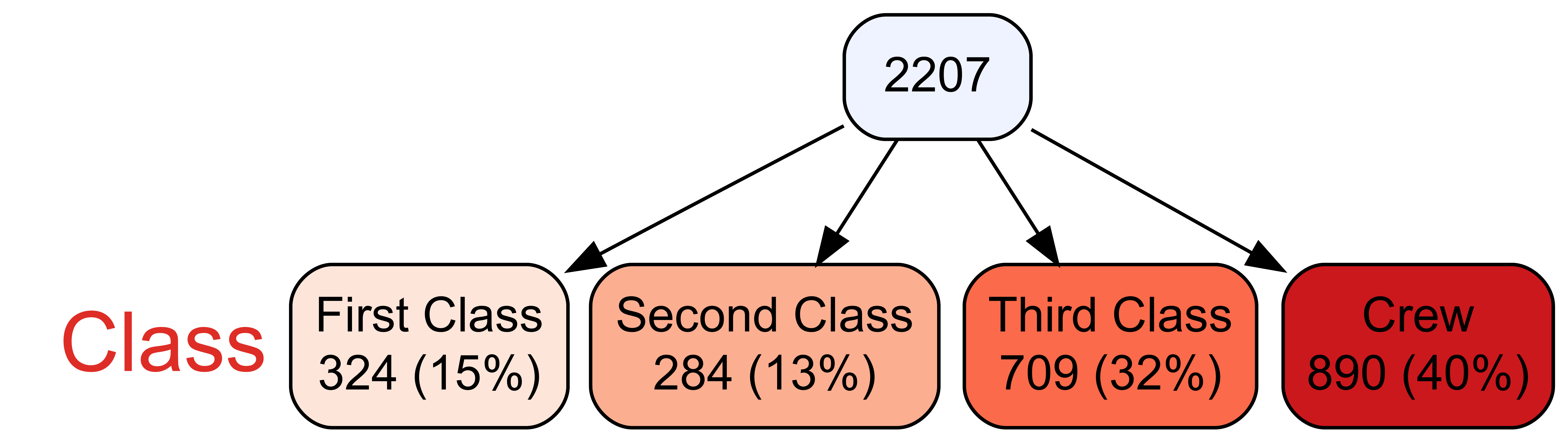} \end{Schunk}
\caption{Using the labelnode parameter.}\label{fig:labelnode}
\end{figure}

\subsection*{Specification of variables} \label{sec:VarSpec}

For convenience, in the call to the \code{vtree} function,
you can specify variable names (separated by whitespace) in a single character string.
(If, however, any of the variable names have internal spaces,
the variable names must be specified as a vector of character strings.)
Additionally, several modifiers can be used, as detailed below.

If an individual variable name is preceded by \code{is.na:},
that variable will be replaced by a missing value indicator in the variable tree.
This facilitates exploration of missing data, for example:

\begin{figure}[H]
\centering
\begin{Schunk}
\begin{Sinput}
R> vtree(td, "Class is.na:fare", horiz = FALSE)
\end{Sinput}

\includegraphics[width=350px]{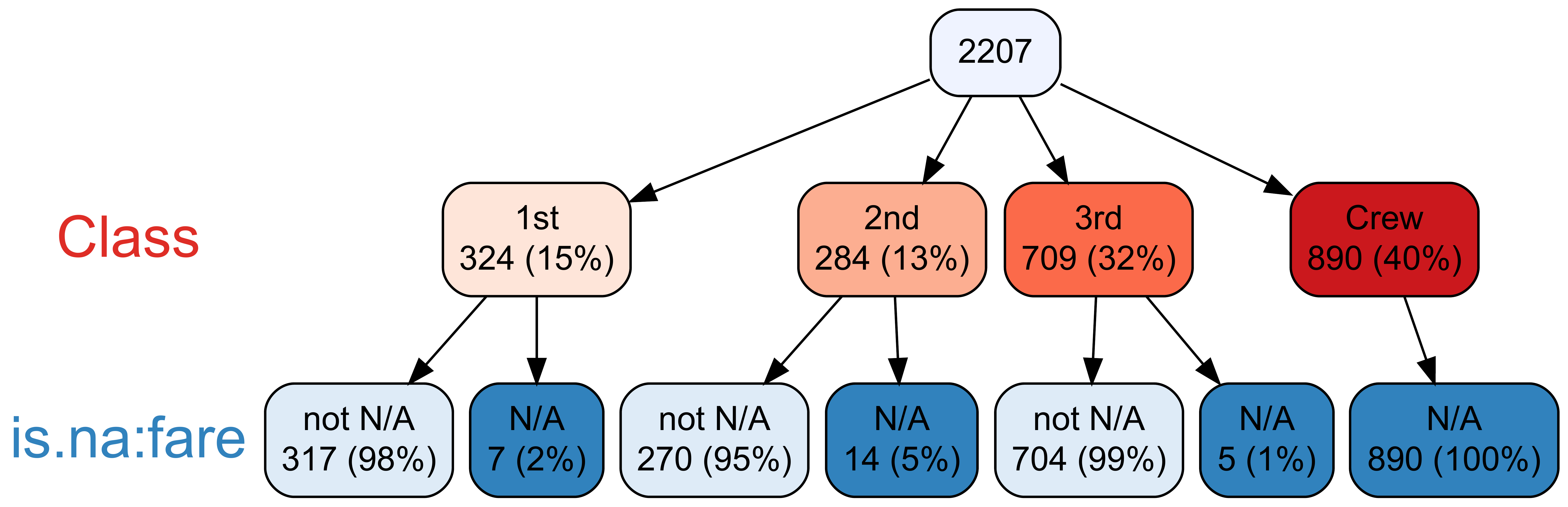} \end{Schunk}
\caption{Using the is.na: variable specification.}\label{fig:isna}
\end{figure}

A variety of other specifications are available.
For example \code{<}, \code{=}, and \code{>} can be used to dichotomize
numeric variables.
While this is a powerful tool for data exploration,
a word of caution is needed.
To ensure scientific rigor,
it is essential that this functionality not be used to explore a variety of
dichotomizations of a predictor variable in relation to the outcome variable.
There is a large literature on the misuse of dichotomization and its
detrimental effect on statistical inference \citep{Altman1994dichotomizing}.
It is therefore recommended that any dichotomization using \pkg{vtree}
be conducted according to a pre-specified protocol \citep{Huebner2016}.

For example, consider the fares paid by passengers (in pounds sterling).
To examine the class of passengers split according to whether they 
paid more than \textsterling 50 or not:

\begin{figure}[H]
\centering
\begin{Schunk}
\begin{Sinput}
R> vtree(td, "fare>50 Class", horiz = FALSE)
\end{Sinput}

\includegraphics[width=400px]{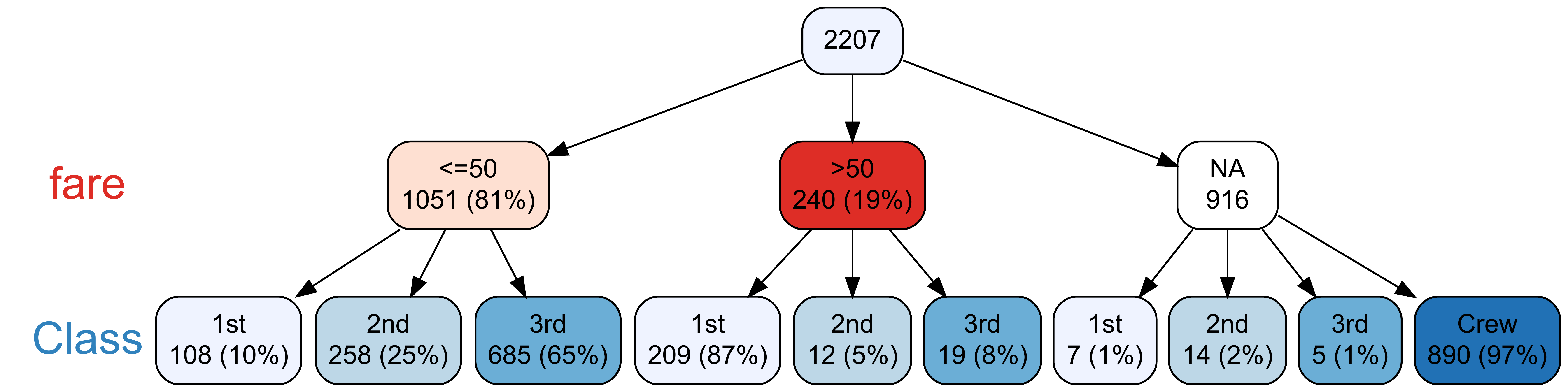} \end{Schunk}
\caption{Using the > variable specification.}\label{fig:dichot}
\end{figure}

\subsection*{Displaying summary statistics in nodes}

It is often useful to display information about \textit{other} variables
(apart from those that define the tree) in the nodes of a variable tree.
This is particularly useful for numeric variables,
which generally would not be used to build the tree since they have too many distinct values.
The \code{summary} parameter allows you to show information
(for example, the mean of a numeric variable)
within each subset of the data frame.

To obtain summary information about fares paid by all of the passengers
on board the Titanic (i.e., in the root node),
you don't need to specify any variables for the tree itself:

\begin{figure}[H]
\centering
\begin{Schunk}
\begin{Sinput}
R> vtree(td, summary = "fare", horiz = FALSE)
\end{Sinput}

\includegraphics[width=100px]{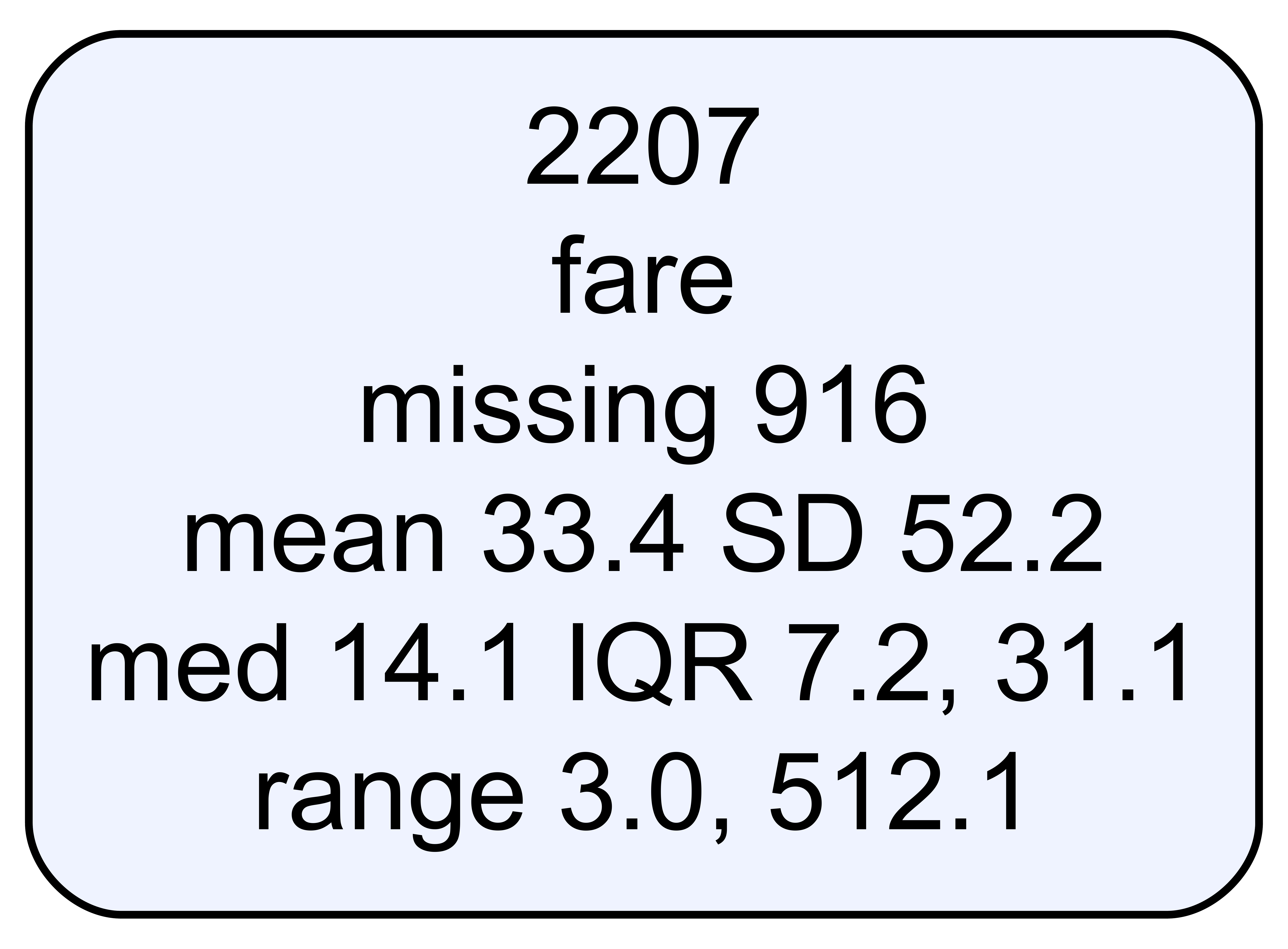} \end{Schunk}
\caption{Summary information over the entire data frame.}\label{fig:summaryroot}
\end{figure}

Using a \textit{summary code}, \code{\%mean\%},
the mean of this variable can be displayed within,
for example, levels of \code{Region} and \code{Class}.
(To reduce the size of this tree we'll hide the crew and the ``Other'' region.)

\begin{figure}[H]
\centering
\begin{Schunk}
\begin{Sinput}
R> vtree(td, "Region Class", summary = "fare \nmean %mean%", horiz = FALSE,
+   prune = list(Region = "Other", Class = "Crew"), splitwidth = 5)
\end{Sinput}

\includegraphics[width=425px]{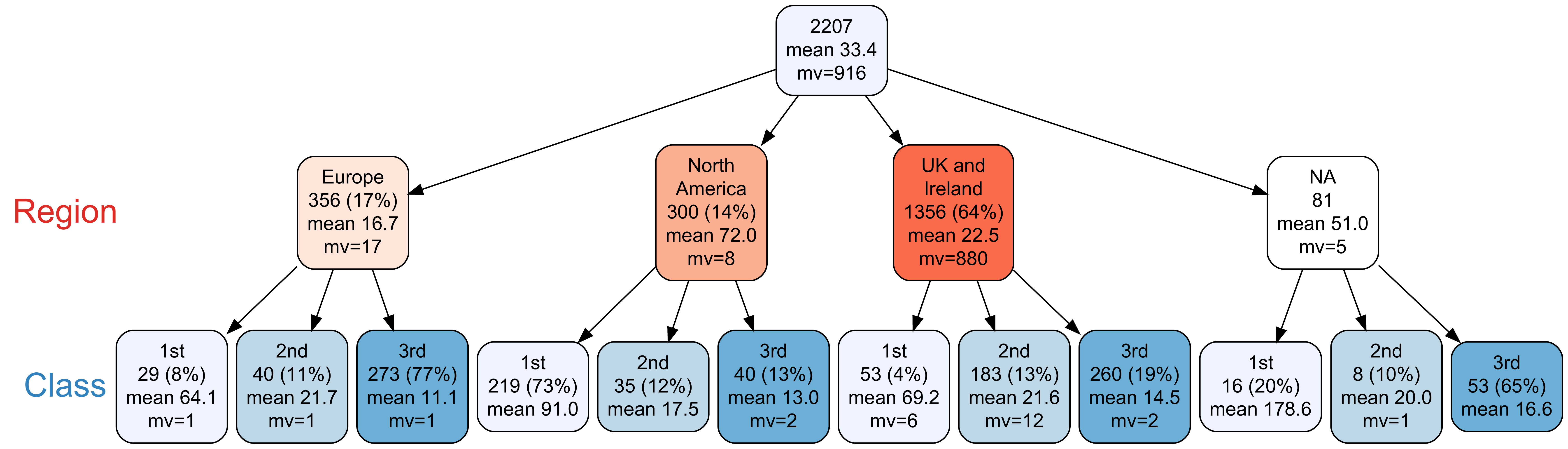} \end{Schunk}
\caption{Using the summary parameter to produce customized summaries.}\label{fig:summarycustom}
\end{figure}

Summary codes always start and end with \code{\%}.
A list is shown in Table~\ref{tab:summaryCodes}.

\begin{table}[H]
\centering
\begin{tabular}{lll}
\hline
\textbf{summary code} & \textbf{variant}  & \textbf{result} \\
\hline
\code{\%mean\%}       & \code{\%meanx\%}*  & mean \\
\code{\%SD\%}         & \code{\%SDx\%}*    & standard deviation  \\
\code{\%sum\%}        & \code{\%sumx\%}*   & sum \\
\code{\%min\%}        & \code{\%minx\%}*   & minimum \\
\code{\%max\%}        & \code{\%maxxx\%}*  & maximum \\
\code{\%range\%}      & \code{\%rangex\%}* & range \\
\code{\%median\%}     & \code{\%median\%}* & median \\
\code{\%IQR\%}        & \code{\%IQRx\%}*   & interquartile range \\
\code{\%freq\%}       & \code{\%freq_\%}** & frequency of values of a variable \\
\code{\%npct\%}       &                    & frequency and percentage \\
\code{\%pct\%}        &                    & same as \code{\%npct\%} but percentage only \\
 \code{\%list\%}      & \code{\%list_\%}** & list of individual values, separated by commas  \\
\hline
\multicolumn{3}{l}{*\textit{Missing values are suppressed. Caution is recommended.}} \\
\multicolumn{3}{l}{**\textit{shows each value on a separate line.}} \\
\end{tabular}
\caption{Summary codes.}\label{tab:summaryCodes}
\end{table}

Sometimes,
you might want to only show summary information in particular nodes.
Table \ref{tab:controlCodes}
lists codes to control where summary information is shown.

\begin{table}[H]
\centering
\begin{tabular}{ll}
\hline
\textbf{code}             & \textbf{summary information restricted to} \\
\hline
\code{\%noroot\%}   & all nodes \textit{except} the root \\
\code{\%leafonly\%} & leaf nodes \\
\code{\%var=v\%}    & nodes of variable v \\
\code{\%node=n\%}   & nodes named n \\
\hline
\end{tabular}

\caption{Control codes.}\label{tab:controlCodes}
\end{table}

\subsection*{Pattern trees}

Each node in a variable tree provides the frequency of a particular combination
of values of the variables.
The leaf nodes represent the observed combinations of values
of \textit{all} of the variables.
For example, in a variable tree of \code{Gender} nested within \code{Class},
the leaf nodes correspond to Male and Female.
These combinations, or \textit{patterns}, can be treated as an additional variable.
And if this new pattern variable is used as the first variable in a tree,
then the branches of the tree will be simplified:
each branch will represent a unique pattern, with no sub-branches.
A \textit{pattern tree} can be easily produced by specifying \code{pattern=TRUE}.
For example:

\begin{figure}[H]
\centering
\begin{Schunk}
\begin{Sinput}
R> vtree(td, "Class Gender", horiz = FALSE, pattern = TRUE)
\end{Sinput}

\includegraphics[width=400px]{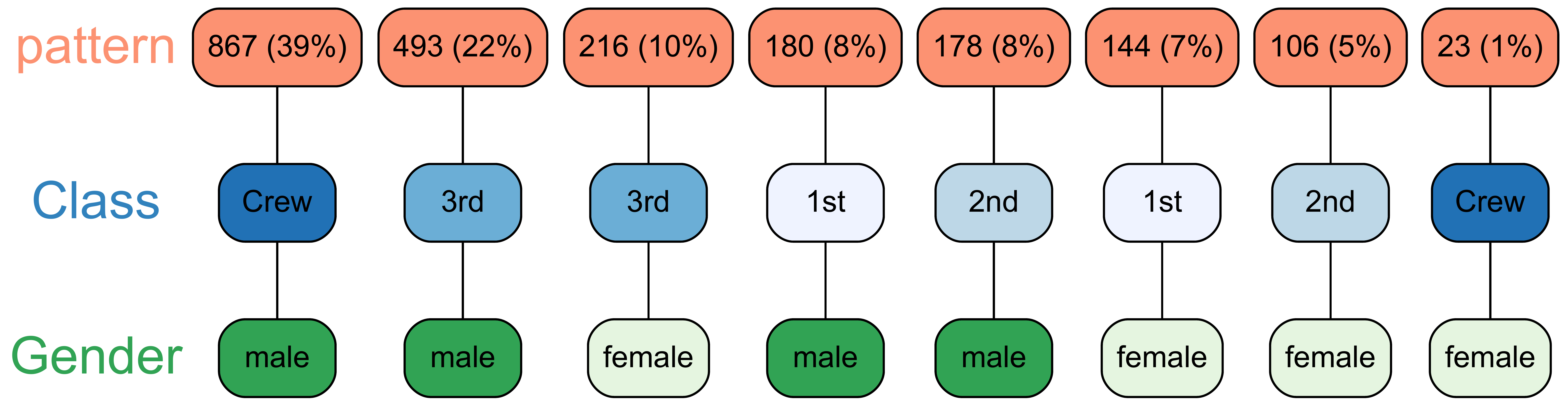} \end{Schunk}
\caption{A pattern tree.}\label{fig:pattern1}
\end{figure}

A special pattern tree with missing values shown as darker nodes
can be produced using the \code{check.is.na} parameter.
Additionally the \code{showlegend} parameter produces marginal frequencies
for each variable.

\begin{figure}[H]
\centering
\begin{Schunk}
\begin{Sinput}
R> vtree(td, "Region Gender age fare", check.is.na = TRUE, 
+   horiz = FALSE, showlegend = TRUE)
\end{Sinput}

\includegraphics[width=375px]{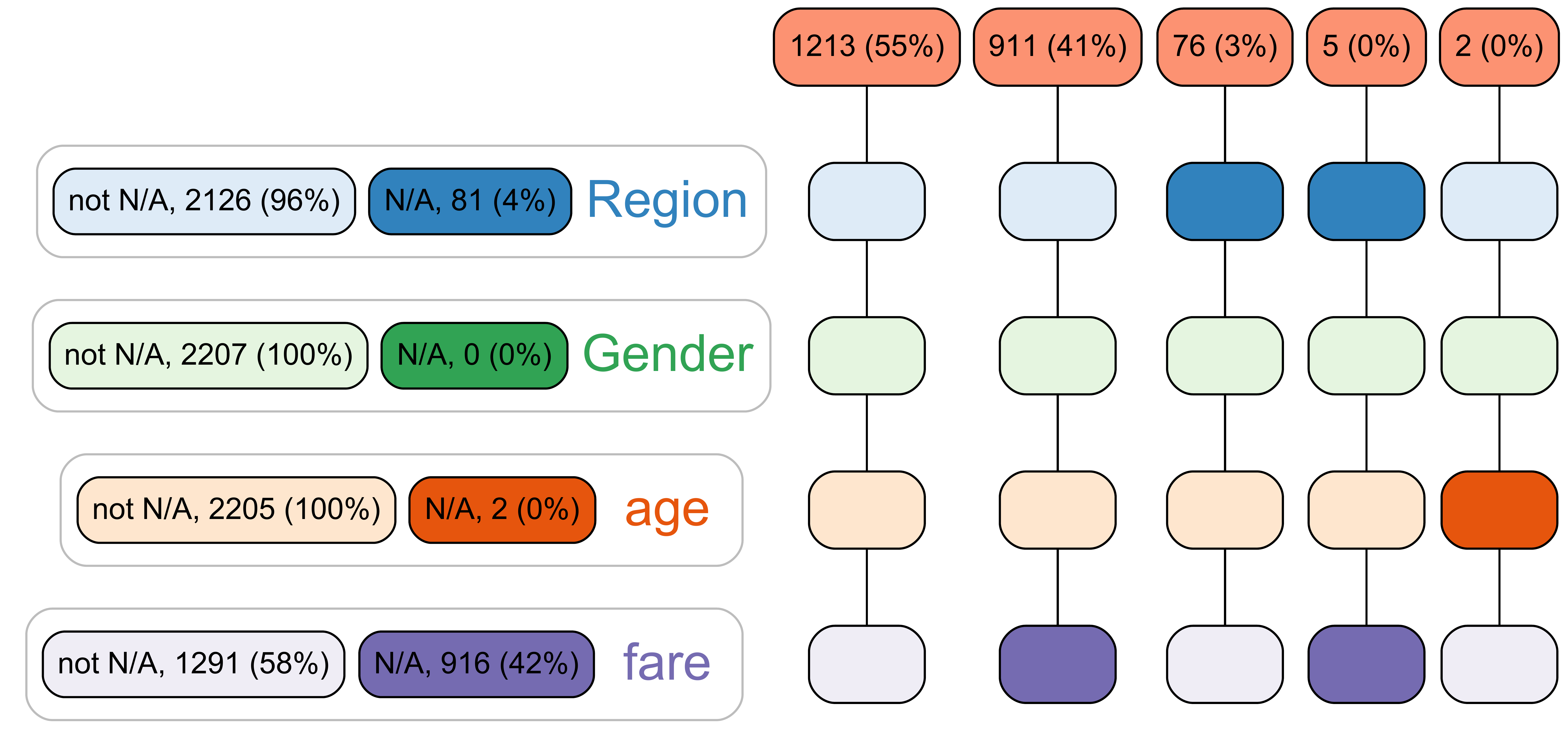} \end{Schunk}
\caption{Using the check.is.na parameter.}\label{fig:checkisna}
\end{figure}

\section{Case Study: A study flow diagram} \label{sec:CONSORTstyle}

Study flow diagrams provide a visual representation of how participants
(or study units) meet or do not meet a sequence of inclusion criteria.
These diagrams provide critical information to the reader of published study.
Medical research in particular has embraced these data visualizations 
as part of recommended reporting guidelines.
Randomized clinical trials use CONSORT diagrams
to show the flow of participants through a single study \citep{CONSORT:2010}.
Systematic reviews use PRISMA flow diagrams to depict study screening \citep{page2020prisma}, \citep{stovold2014study}.
While presenting study flow diagrams is widely considered to be best practice,
preparing these diagrams has traditionally been a
slow, resource intensive, manual process,
which has to be repeated when small changes are made to the data.

\pkg{vtree} uses an \proglang{R} data frame to make a data-driven study flow diagram.
This automates the production of study flow diagrams.
As more data arrives,
data cleaning changes the existing data and
the analysis plan is modified after initial assessment of the data \citep{Huebner2016},
the study flow diagram is easily kept up to date.
Not only does this increase efficiency,
it minimizes the risk of introducing human error.

Consider, for example, the Remdesivir trial of \cite{Spinner:2020},
in which 612 patients 
with confirmed severe acute respiratory syndrome coronavirus 2 (SARS-CoV-2)
infection and moderate COVID-19 pneumonia were screened for inclusion.
Although, in this case, the full data set is not publicly available,
the variables required for the flow diagram can be reconstructed from
Figure~1 of the published paper.
The \code{build.data.frame} function built into the \pkg{vtree} package
makes it easy to construct a data frame indicating
which participants were screened, included
(and of these, who was eligible, and who consented),
the group participants were randomized to, and who started the intervention.
(Additional details have been omitted for the sake of brevity.)

\begin{Schunk}
\begin{Sinput}
R> rem <- build.data.frame(
+   c(   "included","elig","consent","randgrp","started"),
+   list(0,          0,     1,        0,       0,          13),
+   list(0,          1,     0,        0,       0,          3),
+   list(1,          1,     1,        1,       1,          193),
+   list(1,          1,     1,        1,       0,          4),
+   list(1,          1,     1,        2,       1,          191),
+   list(1,          1,     1,        2,       0,          8),
+   list(1,          1,     1,        3,       1,          200))
\end{Sinput}
\end{Schunk}

Next, let's define node labels:

\begin{Schunk}
\begin{Sinput}
R> nodelabels <- list(
+   included = c("Randomized" = "1", "Excluded" = "0"),
+   randgrp=c(
+     "Randomized to receive 10 d of remdesivir" = "1",
+     "Randomized to receive 5 d of remdesivir" = "2",
+     "Randomized to continue standard care" = "3"),
+   started=c(
+     "Did not start remdesivir" = "0",
+     "Started remdesivir" = "1"))
\end{Sinput}
\end{Schunk}

Having set up these objects,
the code to produce a CONSORT-style diagram is fairly straightforward.
In particular, 
the \code{follow} parameter makes it easy to specify which branches
of the tree should be retained.

\begin{figure}[H]
\centering
\begin{Schunk}
\begin{Sinput}
R> vtree(rem,"included randgrp started",
+   labelnode = nodelabels,
+   follow = list(included = "1", randgrp = c("1", "2")),
+   summary = c(
+     "consent=0 \n(Withdrew consent %sum%%var=included%%node=0%)",
+     "elig=0 \n(Ineligible %sum%%var=included%%node=0%)"),
+   cdigits = 0, showvarnames = FALSE, title = "patients screened",
+   horiz = FALSE, fillcolor = "lightsteelblue1", showpct = FALSE)
\end{Sinput}

\includegraphics[width=400px]{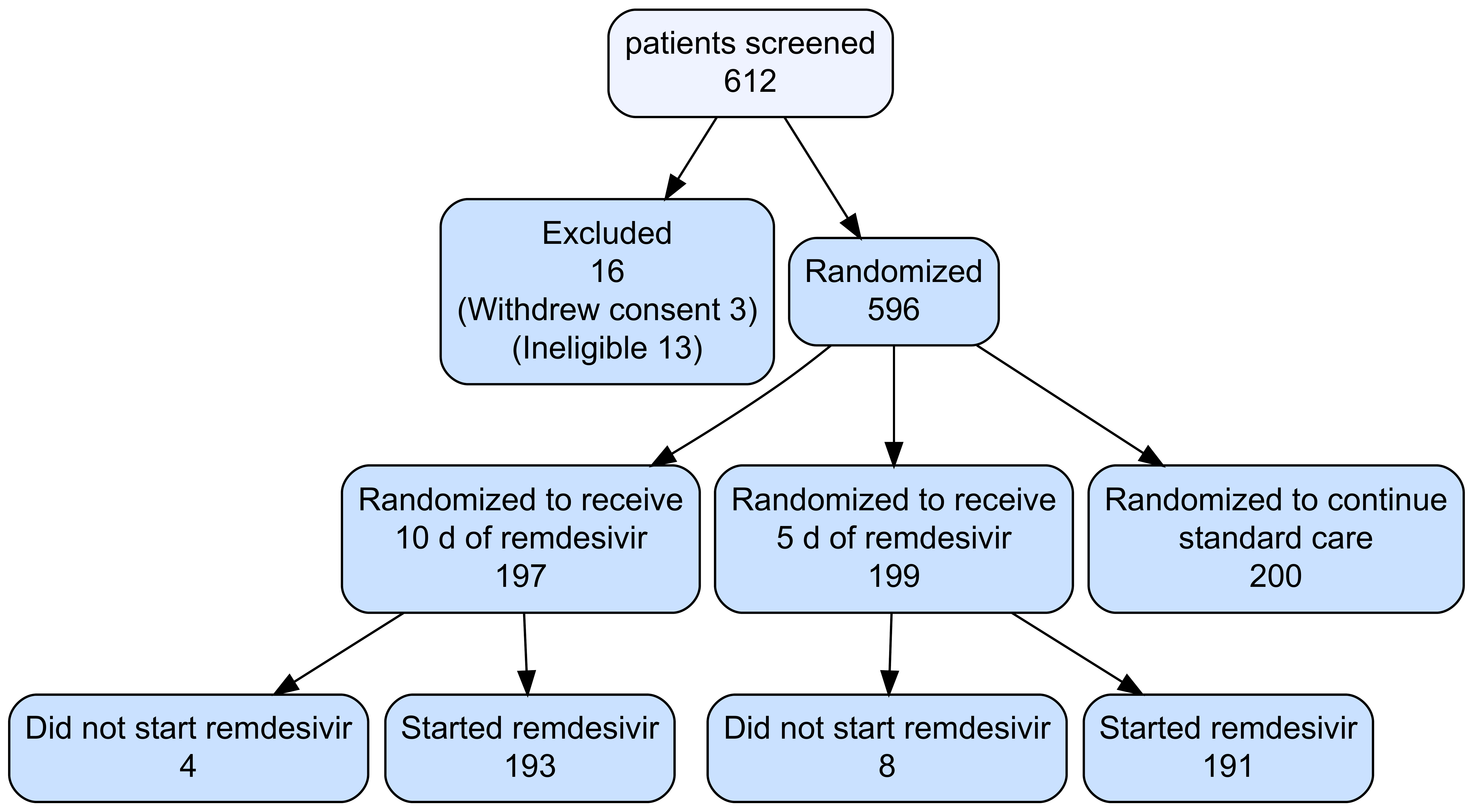} \end{Schunk}
\caption{A variable tree providing a CONSORT-style diagram for the Remdesivir trial.}\label{fig:CONSORTstyle}
\end{figure}

\section{Case Study: Ottawa Police Service Traffic Stops Data} \label{sec:TrafficStops}

Following a 2005 racial profiling complaint to the Ontario Human Rights Commission,
the Ottawa Police Service agreed to collect race data in traffic stops,
known as the Traffic Stop Race Data Collection Project (TSRDCP).
The TSRDCP required police officers to record \textit{their perception} of the driver’s race, for traffic stops over a two-year period from June 27, 2013 to June 26, 2015.
A data set representing these traffic stops was made public
(\url{https://www.ottawapolice.ca/en/news-and-community/race-archive.aspx}).

Important questions concern whether some racialized or ethnic groups are
stopped at a rate disproportionate to overall makeup of the population. 
This would require external data, not presented here.
See the report by researchers at York University, dated October 2016,
for a comprehensive analysis:
\url{https://www.ottawapolice.ca/en/about-us/resources/.TSRDCP_York_Research_Report.pdf}

In the York University report,
some records from the raw data were removed due to errors.
Additionally, since some drivers were stopped more than once,
only a single report per driver was included.
It was not possible to replicate this last step because driver identifiers
were not included in the publicly available data set.

\begin{figure}[H]
\centering
\begin{Schunk}
\begin{Sinput}
R> vtree(z, "race=white age gender", splitwidth = Inf, sameline = TRUE,
+   summary = "how_cleared=Final \nNo action %pct%",
+   title = "Total stops", showlegend = TRUE, showlegendsum = TRUE)
\end{Sinput}

\includegraphics[width=425px]{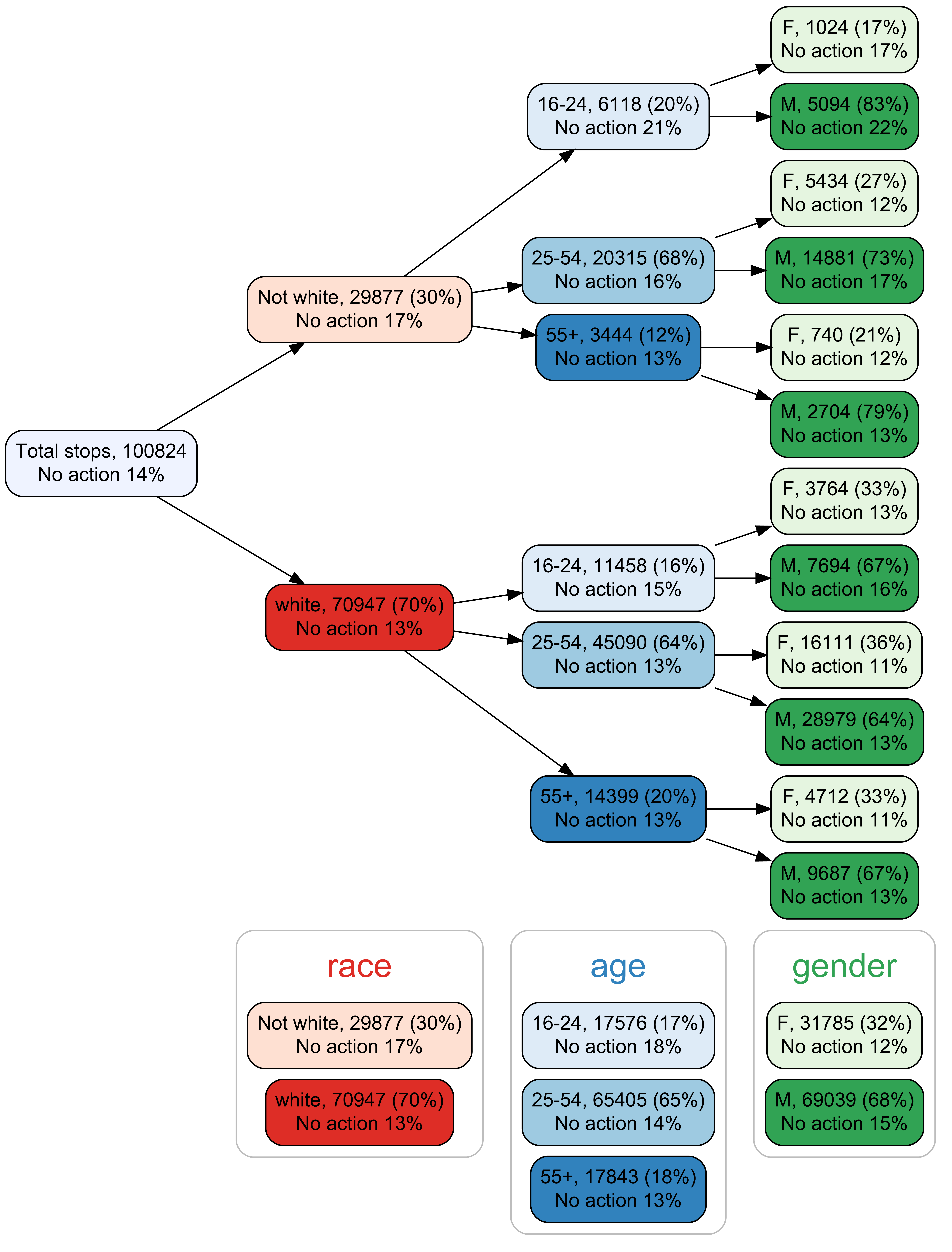} \end{Schunk}
\caption{Variable tree for \{\code{race=white}\} $\rightarrow$ \code{age} $\rightarrow$ \code{gender}. Each node also shows the percentage of traffic stops with ``final/no action'' outcome.}\label{fig:FinalOutcome}
\end{figure}

The data set includes a number of variables including age group and
gender of the driver.
One important variable is the outcome (\code{how_cleared}) of the traffic stop:
\textit{charged}, \textit{warning}, or \textit{final (no action)}.
This last outcome is of particular interest,
because it means that the driver was neither charged nor given a warning,
which may raise the question of whether the stop was actually necessary.
Figure~\ref{fig:FinalOutcome} shows the percentage of stops
with this outcome (via the \code{summary} parameter) in each node of a tree for
\{\code{race=white}\} $\rightarrow$ \code{age} $\rightarrow$ \code{gender}.
Here race has been dichotomized as white or non-white, denoted \{\code{race=white}\}
(where the braces are for clarity in the arrow notation).
Legend nodes are shown for each variable.
Additionally the percentage of stops for which no action was taken is also
shown in the legend nodes because \code{showlegendsum = TRUE} has been specified.

A number of interesting patterns emerge.
The following drivers were more likely to receive neither a charge nor a warning:
(1) male drivers, within all combinations of race and age;
(2) younger drivers, within each race; and
(3) non-white drivers.

\newpage

\section{Concluding remarks}

Variable trees are an intuitive way to represent discrete multivariate data.
The \pkg{vtree} package in \proglang{R}
provides an implementation of variable trees
along with a number of convenient extensions.
There are a variety of other methods for displaying discrete multivariate data,
and depending on the context, one of these methods may be preferable.
However, the simple structure of variable trees provides
not only ease of interpretation but also considerable generality.
We have found that variable trees facilitate iterative data exploration 
when a statistician is working together with a domain expert.

A key characteristic of variable trees is that the order of variables is important.
Sometimes the ordering of variables is natural
(e.g., school board $\rightarrow$ school $\rightarrow$ teacher),
in other cases it is dictated by the research question,
and in still other cases the choice of ordering is up to the analyst.
Depending on the situation,
the variable-order dependence may be a strength or a weakness.

While \pkg{vtree} can be used to explore data,
it can also be used to generate study flow diagrams.
In recent years there has been growing concern about the ``reproducibility crisis'' in science
\citep{BakerReproducibility}.
In order to produce study flow diagrams using \pkg{vtree},
all of the variables and the corresponding set of inclusion/exclusion steps
must be in a single data frame, which encourages a reproducible workflow.
The design of \pkg{vtree} was influenced by the
tidyverse philosophy \citep{Wickham2019tidyverse},
with its emphasis on reproducible workflows,
and \pkg{vtree} works well with tidyverse tools.
A key barrier to the wider adoption of study flow diagrams has been 
the difficulty required to produce them.
\pkg{vtree} facilitates reproducible research by making it easy to produce accurate study flow diagrams directly from the study data.

To conclude,
variable trees are an intuitive new data exploration tool for visualizing nested subsets.
Applications of variable trees include revealing patterns in data,
understanding missingness and producing study flow diagrams for reproducible research.

\section*{Acknowledgments}

The \pkg{vtree} package builds on the \pkg{DiagrammeR} package \citep{DiagrammeRpackage},
which in turn is based on the
\textit{Graphviz} graph visualization software \citep{Gansner00anopen}.
Sebastian Gatscha also contributed code to \pkg{vtree}.
Development of \pkg{vtree} was partially supported by
the Clinical Research Unit (CRU) 
at the Children's Hospital of Eastern Ontario Research Institute.
Members of the CRU contributed helpful suggestions and endless patience.

\bibliography{refs}

\begin{thebibliography}{29}
\newcommand{\enquote}[1]{``#1''}
\providecommand{\natexlab}[1]{#1}
\providecommand{\url}[1]{\texttt{#1}}
\providecommand{\urlprefix}{URL }
\expandafter\ifx\csname urlstyle\endcsname\relax
  \providecommand{\doi}[1]{doi:\discretionary{}{}{}#1}\else
  \providecommand{\doi}{doi:\discretionary{}{}{}\begingroup
  \urlstyle{rm}\Url}\fi
\providecommand{\eprint}[2][]{\url{#2}}

\bibitem[{Altman(1994)}]{Altman1994dichotomizing}
Altman DG (1994).
\newblock \enquote{Problems in Dichotomizing Continuous Variables.}
\newblock \emph{American Journal of Epidemiology}, \textbf{139}(4), 442--442.

\bibitem[{Baker(2016)}]{BakerReproducibility}
Baker M (2016).
\newblock \enquote{1,500 Scientists Lift the Lid on Reproducibility.}
\newblock \emph{Nature}, \textbf{533}(7604), 452--454.

\bibitem[{Bangdiwala and Shankar(2013)}]{bangdiwala2013agreement}
Bangdiwala SI, Shankar V (2013).
\newblock \enquote{The agreement chart.}
\newblock \emph{BMC medical research methodology}, \textbf{13}(1), 1--7.

\bibitem[{Chen(2018)}]{VennDiagramPackage}
Chen H (2018).
\newblock \emph{VennDiagram: Generate High-Resolution Venn and Euler Plots}.
\newblock R package version 1.6.20,
  \urlprefix\url{https://CRAN.R-project.org/package=VennDiagram}.

\bibitem[{Conway \emph{et~al.}(2017)Conway, Lex, and Gehlenborg}]{Conway:2017}
Conway J, Lex A, Gehlenborg N (2017).
\newblock \enquote{UpSetR: An R Package for the Visualization of Intersecting
  Sets and their Properties.}
\newblock \emph{Bioinformatics (Oxford, England)}, \textbf{33}.
\newblock \doi{10.1093/bioinformatics/btx364}.

\bibitem[{Gansner and North(2000)}]{Gansner00anopen}
Gansner ER, North SC (2000).
\newblock \enquote{An Open Graph Visualization System and its Applications to
  Software Engineering.}
\newblock \emph{Software: Practice and Experience}, \textbf{30}(11),
  1203--1233.

\bibitem[{Hartigan and Kleiner(1981)}]{HartiganKleiner:1981}
Hartigan J, Kleiner B (1981).
\newblock \enquote{Mosaics for Contingency Tables.}
\newblock \emph{Computer Science and Statistics: Proceedings of the 13th
  Symposium on the Interface}.
\newblock \doi{10.1007/978-1-4613-9464-8_37}.

\bibitem[{Hellerstein \emph{et~al.}(2017)}]{Hellerstein2017PrinciplesOD}
Hellerstein JM, \emph{et~al.} (2017).
\newblock \emph{Principles of Data Wrangling Practical Techniques for Data
  Preparation}.
\newblock O'Reilly Media, Inc., Sebastopol, CA, USA.

\bibitem[{Hornik \emph{et~al.}(2006)Hornik, Zeileis, and
  Meyer}]{hornik2006strucplot}
Hornik K, Zeileis A, Meyer D (2006).
\newblock \enquote{The strucplot Framework: Visualizing Multi-Way Contingency
  Tables with vcd.}
\newblock \emph{Journal of Statistical Software}, \textbf{17}(3), 1--48.

\bibitem[{Huebner \emph{et~al.}(2016)}]{Huebner2016}
Huebner M, \emph{et~al.} (2016).
\newblock \enquote{A Systematic Approach to Initial Data Analysis is Good
  Research Practice.}
\newblock \emph{The Journal of Thoracic and Cardiovascular Surgery},
  \textbf{151}(1), 25--27.
\newblock ISSN 0022-5223.
\newblock \doi{10.1016/j.jtcvs.2015.09.085}.
\newblock \urlprefix\url{https://doi.org/10.1016/j.jtcvs.2015.09.085}.

\bibitem[{Iannone(2020)}]{DiagrammeRpackage}
Iannone R (2020).
\newblock \emph{DiagrammeR: Graph/Network Visualization}.
\newblock R package version 1.0.6.1,
  \urlprefix\url{https://CRAN.R-project.org/package=DiagrammeR}.

\bibitem[{Jeppson \emph{et~al.}(2020)Jeppson, Hofmann, and
  Cook}]{ggmosaicPackage}
Jeppson H, Hofmann H, Cook D (2020).
\newblock \emph{ggmosaic: Mosaic Plots in the 'ggplot2' Framework}.
\newblock R package version 0.3.0,
  \urlprefix\url{http://github.com/haleyjeppson/ggmosaic}.

\bibitem[{Kassambara(2020)}]{ggpubrPackage}
Kassambara A (2020).
\newblock \emph{ggpubr: 'ggplot2' Based Publication Ready Plots}.
\newblock R package version 0.4.0,
  \urlprefix\url{https://CRAN.R-project.org/package=ggpubr}.

\bibitem[{{Katal} \emph{et~al.}(2013){Katal}, {Wazid}, and {Goudar}}]{BigData}
{Katal} A, {Wazid} M, {Goudar} RH (2013).
\newblock \enquote{Big Data: Issues, Challenges, Tools and Good Practices.}
\newblock In \emph{2013 Sixth International Conference on Contemporary
  Computing (IC3)}, pp. 404--409.

\bibitem[{Larsson(2020)}]{eurlerrPackage}
Larsson J (2020).
\newblock \emph{{eulerr}: Area-Proportional {Euler} and {Venn} Diagrams with
  Ellipses}.
\newblock R package version 6.1.0,
  \urlprefix\url{https://cran.r-project.org/package=eulerr}.

\bibitem[{Lex \emph{et~al.}(2014)}]{Lex:2014}
Lex, \emph{et~al.} (2014).
\newblock \enquote{UpSet: Visualization of Intersecting Sets.}
\newblock \emph{IEEE Transactions on Visualization and Computer Graphics},
  \textbf{20}, 1983--1992.
\newblock \doi{10.1109/TVCG.2014.2346248}.

\bibitem[{Moon(2017)}]{moon2017learn}
Moon KW (2017).
\newblock \emph{Learn ggplot2 using shiny App}, chapter Balloon Plot.
\newblock Springer.

\bibitem[{Nelson and Rafferty(2012)}]{Nelson2012Nightingale}
Nelson S, Rafferty AM (2012).
\newblock \emph{Notes on Nightingale: The Influence and Legacy of a Nursing
  Icon}.
\newblock Cornell University Press.

\bibitem[{Page \emph{et~al.}(2020)}]{page2020prisma}
Page MJ, \emph{et~al.} (2020).
\newblock \enquote{The PRISMA 2020 Statement: An Updated Guideline for
  Reporting Systematic Reviews.}
\newblock \doi{10.31222/osf.io/v7gm2}.
\newblock \urlprefix\url{osf.io/preprints/metaarxiv/v7gm2}.

\bibitem[{Philipp \emph{et~al.}(2016)Philipp, Zeileis, and
  Strobl}]{stablelearnerPackage}
Philipp M, Zeileis A, Strobl C (2016).
\newblock \enquote{A Toolkit for Stability Assessment of Tree-Based Learners.}
\newblock In A~Colubi, A~Blanco, C~Gatu (eds.), \emph{Proceedings of {COMPSTAT}
  2016 -- 22nd International Conference on Computational Statistics}, pp.
  315--325. The International Statistical Institute/International Association
  for Statistical Computing.
\newblock ISBN 978-90-73592-36-0.
\newblock Preprint available at
  \url{http://EconPapers.RePEc.org/RePEc:inn:wpaper:2016-11}.

\bibitem[{Schulz \emph{et~al.}(2010)Schulz, Altman, and Moher}]{CONSORT:2010}
Schulz KF, Altman DG, Moher D (2010).
\newblock \enquote{CONSORT 2010 Statement: Updated Guidelines for Reporting
  Parallel Group Randomised Trials.}
\newblock \emph{BMJ}, \textbf{340}.
\newblock ISSN 0959-8138.
\newblock \doi{10.1136/bmj.c332}.
\newblock \eprint{https://www.bmj.com/content},
  \urlprefix\url{https://www.bmj.com/content/340/bmj.c332}.

\bibitem[{Spinner \emph{et~al.}(2020)}]{Spinner:2020}
Spinner CD, \emph{et~al.} (2020).
\newblock \enquote{{Effect of Remdesivir vs Standard Care on Clinical Status at
  11 Days in Patients With Moderate COVID-19: A Randomized Clinical Trial}.}
\newblock \emph{JAMA}.
\newblock ISSN 0098-7484.
\newblock \doi{10.1001/jama.2020.16349}.
\newblock
  \eprint{https://jamanetwork.com/journals/jama/articlepdf/2769871/jama\_spinner\_2020\_oi\_200097\_1597942780.67456.pdf},
  \urlprefix\url{https://doi.org/10.1001/jama.2020.16349}.

\bibitem[{Stovold \emph{et~al.}(2014)}]{stovold2014study}
Stovold E, \emph{et~al.} (2014).
\newblock \enquote{Study Flow Diagrams in Cochrane Systematic Review Updates:
  An Adapted Prisma Flow Diagram.}
\newblock \emph{Systematic reviews}, \textbf{3}(1), 54.

\bibitem[{Urbanek and Wichtrey(2018)}]{iplotsPackage}
Urbanek S, Wichtrey T (2018).
\newblock \emph{iplots: Interactive Graphics for R}.
\newblock R package version 1.1-7.1,
  \urlprefix\url{https://CRAN.R-project.org/package=iplots}.

\bibitem[{Waldrop(2016)}]{waldrop2016chips}
Waldrop MM (2016).
\newblock \enquote{The Chips are Down for Moore's Law.}
\newblock \emph{Nature News}, \textbf{530}(7589), 144.

\bibitem[{Wickham \emph{et~al.}(2019)}]{Wickham2019tidyverse}
Wickham H, \emph{et~al.} (2019).
\newblock \enquote{Welcome to the Tidyverse.}
\newblock \emph{Journal of Open Source Software}, \textbf{4}(43), 1686.

\bibitem[{Wilkinson(2011)}]{venneulerPackage}
Wilkinson L (2011).
\newblock \emph{venneuler: Venn and Euler Diagrams}.
\newblock R package version 1.1-0,
  \urlprefix\url{https://CRAN.R-project.org/package=venneuler}.

\bibitem[{{Wilkinson}(2012)}]{Wilkinson:2012}
{Wilkinson} L (2012).
\newblock \enquote{Exact and Approximate Area-Proportional Circular Venn and
  Euler Diagrams.}
\newblock \emph{IEEE Transactions on Visualization and Computer Graphics},
  \textbf{18}(2), 321--331.

\bibitem[{Zeileis \emph{et~al.}(2007)Zeileis, Meyer, and
  Hornik}]{zeileis2007residual}
Zeileis A, Meyer D, Hornik K (2007).
\newblock \enquote{Residual-Based Shadings for Visualizing (Conditional)
  Independence.}
\newblock \emph{Journal of Computational and Graphical Statistics},
  \textbf{16}(3), 507--525.

\end{thebibliography}

\end{document}